\documentclass[aps,prb,twocolumn,floats,floatfix,epsfig,pdflatex]{revtex4}
\usepackage{amssymb,amsbsy,amsmath,mathrsfs}
\usepackage{epsfig,psfrag,subfigure}
\usepackage{graphicx,float}
\usepackage{times}
\usepackage{color}
\usepackage{subfigure}
\usepackage{setspace}
\usepackage{bm} 

\newcommand\bea{\begin{eqnarray}}
\newcommand\eea{\end{eqnarray}}
\newcommand\beq{\begin{equation}}
\newcommand\eeq{\end{equation}}
\newcommand\bib{\bibitem}

\newcommand{\noi}{\noindent}
\newcommand{\non}{\nonumber}
\newcommand{\al}{\alpha}
\newcommand{\de}{\delta}

\newcommand{\ga}{\gamma}

\newcommand{\si}{\sigma}
\newcommand{\ta}{\theta}
\newcommand{\om}{\omega}

\newcommand{\la}{\langle}
\newcommand{\ra}{\rangle}

\newcommand{\bra}[1]{\langle #1|}
\newcommand{\ket}[1]{|#1\rangle}

\begin{document}

\title{Periodically driven model with quasiperiodic potential and staggered 
hopping amplitudes: engineering of mobility gaps and multifractal states}

\author{Sreemayee Aditya$^1$, K. Sengupta$^2$ and Diptiman Sen$^1$}

\affiliation{$^1$Centre for High Energy Physics, Indian Institute of 
Science, Bengaluru 560012, India \\
$^2$School of Physical Sciences, Indian Association for the 
Cultivation of Science, Jadavpur, Kolkata 700032, India}

\begin{abstract}
We study if periodic driving of a model with a quasiperiodic potential 
can generate interesting Floquet phases which have no counterparts in the 
static model. Specifically, we consider the Aubry-Andr\'e model which is a 
one-dimensional time-independent model with an on-site quasiperiodic potential
$V_0$ and a nearest-neighbor hopping amplitude which is taken to have a 
staggered form. We add a uniform hopping amplitude which varies in time 
either sinusoidally or as a square pulse with 
a frequency $\omega$. Unlike the static Aubry-Andr\'e model which has a 
simple phase diagram with only two phases (only extended or only
localized states), we find that the driven model has four possible phases
for the Floquet eigenstates: a phase with gapless quasienergy bands and only
extended states, a phase with multiple mobility gaps separating different 
quasienergy bands, a mixed phase with coexisting 
extended, multifractal, and localized states, and a phase with only 
localized states. The multifractal states have generalized inverse
participation ratios which scale with the system size with exponents which 
are different from the values for both extended and localized states.
In addition, we observe intricate re-entrant 
transitions between the different kinds of states when $\omega$ and $V_0$
are varied. The appearance of such transitions is confirmed by the behavior 
of the Shannon entropy. Many of our numerical results can be understood from 
an analytic Floquet Hamiltonian derived using a Floquet perturbation theory
which uses the inverse of the driving amplitude as the perturbation parameter. 
In the limit of high frequency and large driving amplitude, 
we find that the Floquet quasienergies match the energies 
of the undriven system, but the Floquet eigenstates are much more extended. We 
also study the spreading of a one-particle wave packet and find that it is 
always ballistic but the ballistic velocity varies significantly with the 
system parameters, sometimes showing a non-monotonic dependence on $V_0$
which does not occur in the static model.
Finally, we compare the results for the driven model with a static staggered 
hopping amplitude with a model with a static uniform hopping amplitudes, 
and we find some significant differences between the two cases. All our results
are found to be independent of the driving protocol, either sinusoidal or
square pulse. We conclude that the interplay of quasiperiodic potential and 
driving produces a rich phase diagram which does not appear in the static model.
\end{abstract}

\maketitle

\section{Introduction}
\label{sec1}

Periodically driven systems have been studied extensively for the 
last several
years because of the multitude of unusual phenomena that they can 
exhibit~\cite{rev1,rev2,rev3,rev4,rev5,rev6,rev7,rev8} which have no
analogs in time-independent systems. For instance, 
periodic driving can be used to engineer topological states of 
matter~\cite{topo1,topo2,topo3,topo4,topo5,topo6}, Floquet time 
crystals~\cite{tc1,tc2,tc3} and other novel steady 
states~\cite{russomanno,lazarides}, produce dynamical
localization~\cite{dynloc1,dynloc2,dynloc3,dynloc4}, dynamical 
freezing~\cite{dynfreez1,dynfreez2,dynfreez3} and other dynamical 
transitions~\cite{heyl1,nandy,heyl2,karrasch,kriel,sirker,canovi,dutta,
sarkar,arze,aditya}, tune a system into ergodic or non-ergodic
phases~\cite{erg1,erg2,erg3}, and generate emergent conservation 
laws~\cite{haldar}.

The effects of periodic driving on localization-delocalization transitions
have been relatively less studied~\cite{sarkar1,sarkar2}. A well-studied 
{\it time-independent} model in one dimension model with a 
localization-delocalization transition is the 
Aubry-Andr\'e model~\cite{aubry,soko}. This is a tight-binding model with
uniform nearest-neighbor hopping amplitude $\ga$ and a quasiperiodic 
on-site potential with strength $V_0$. 
Depending on the value of $V_0$, this
model is known to have two phases, with all states being extended 
(localized) for $V_0$ smaller (larger) than some critical value;
the critical value of $V_0$ is known to be equal to $2 \ga$. 
Studies of other variants of this system have shown that multiple 
localization transitions and states with multifractal properties can 
appear~\cite{li,mishra}. Given these different possibilities for an
undriven system, it would be interesting to study what can happen if 
a system with both a quasiperiodic potential and a staggered hopping
amplitude is driven periodically in time. 
In particular, we would like to understand if the driving can generate 
mobility edges or gaps, and whether states which 
are neither extended nor localized can appear~\cite{roy}. (See 
Ref.~\onlinecite{grover} for a recent experiment on a kicked system 
with a quasiperiodic potential).

The plan of this paper is as follows. In Sec.~\ref{sec2} we present our
model which is essentially the Aubry-Andr\'e model driven periodically 
in a particular way. The Hamiltonian consists of a staggered time-independent 
hopping with values $\ga_1$ and $\ga_2$ on alternating sites, and a 
uniform time-periodic hopping with driving amplitude $a$ and 
frequency $\om$ (both hopping amplitudes are 
between nearest neighbors only), and a quasiperiodic on-site potential with
strength $V_0$. In the limit where both $a$ and $\om$ are much larger 
than the time-independent hopping amplitudes and $V_0$ 
(i.e., $a, ~\om \gg \ga_1, ~\ga_2, ~V_0$), we use a Floquet perturbation 
theory to analytically derive an effective 
time-independent Hamiltonian $H_F$ called the Floquet Hamiltonian. Both 
sinusoidal driving and driving by a periodic square pulse are considered.
In Sec.~\ref{sec3} we numerically study the properties of the eigenstates
of the Floquet operator $U$ (in particular, their scaling with the system
size $L$ to see if they are localized, extended or multifractal), and 
whether there are any mobility edges or gaps in the Floquet quasienergies. 
We compare the results obtained
numerically from the Floquet operator with those obtained from the Floquet
Hamiltonian. We also study the Shannon entropy
and the structure of the Floquet Hamiltonian in real space to shed
more light on the degree of localization of the Floquet eigenstates. 
The form of the Floquet eigenstates and eigenvalues in the high-frequency 
limit is studied. In Sec.~\ref{sec4} we consider the spreading of a wave 
packet and compare the cases where the time-independent hopping amplitudes 
are uniform ($\ga_1 = \ga_2$) and staggered ($\ga_2 = - \ga_1$) respectively.
We use van Vleck perturbation theory to gain an understanding of the 
results for the case of uniform hopping.
In Sec.~\ref{sec5} we summarize our results and point out some directions for
future studies.

Our main results are as follows. We find that in contrast to the 
Aubry-Andr\'e model which has only extended or only localized states
depending on the value of $V_0$, the periodically driven model has
four possible phases: a phase with gapless quasienergy bands and only
extended states, a phase with multiple mobility gaps separating different 
quasienergy bands, a mixed phase with coexisting extended, multifractal, and
localized states, and a phase with only localized states. The multifractal 
states appear at intermediate driving frequencies $\om$ and large values of 
$V_0$; these are eigenstates of the Floquet operator which are neither 
extended nor localized. The multifractal nature is inferred from the 
scaling with the system size of the generalized inverse participation ratio 
of these states; the scaling exponents are found to be different from the
values of both extended and localized states. We find multiple re-entrant
transitions between the different kinds of states for certain ranges of
values of $\om$ and $V_0$; such transitions have no counterparts in the Aubry-Andr\'e model. All this is found to be true 
for both sinusoidal driving and square pulse driving. 
In the high-frequency limit, we find that the spectrum of Floquet quasienergies
approaches the energy spectrum of the model with no driving. However, 
when the driving amplitude is also large, we find that the Floquet eigenstates 
are significantly different from those of the undriven system;
typically, driving makes the states much more extended. We find that
one-particle wave packets always spread ballistically but the ballistic
velocity varies considerably depending on the system parameters. 
Interestingly, for the case of uniform time-independent hopping, we 
sometimes find a non-monotonic dependence of the ballistic velocity on $V_0$.
In conclusion, we find that a combination of quasiperiodic potential, periodic
driving of the uniform hopping amplitude and a time-independent staggered hopping hopping gives rise to a wide range of unusual properties 
of the Floquet eigenstates.

\section{Model Hamiltonian and Floquet perturbation theory}
\label{sec2}

In this section, we will study a one-dimensional model with both uniform 
and staggered nearest-neighbor hopping amplitudes and a quasiperiodic on-site 
potential. We will first consider what happens when the uniform hopping
amplitude is driven sinusoidally in time with a frequency $\om$ and an 
amplitude $a$. Later we will study what happens when the driving is 
given by a periodic square pulse. The Hamiltonian of the system is given by
\bea H (t) &=&\sum_{j} ~[(a\sin (\om t)+\ga_{1}) ~(a_{j}^{\dagger}b_{j} + 
b_j^\dagger a_j) \non \\
&& ~~~~~~+ (a\sin (\om t)+\ga_{2}) ~(a_{j}^{\dagger} b_{j-1} +
b_{j-1}^\dagger a_j) \label{ham1} \\
&& ~~~~~~+ V_{0} ~\{ \cos(2\pi\beta (2j-1)) ~a_{j}^{\dagger}a_{j} \non \\
&& ~~~~~~~~~~~~~~~~~+ \cos(2\pi \beta(2j)) ~b_{j}^{\dagger}b_{j} \}], \non
\eea
where the quasiperiodic potential has strength $V_0$, and we take $\beta =
(\sqrt{5} - 1)/2$ so that it is irrational. We have taken the unit cells,
labeled by an integer $j$, to consist of two sites labeled as $a_j$ and
$b_j$. We will set the nearest-neighbor spacing to be equal to 1; hence
the size of a unit cell is 2. We will impose periodic boundary conditions
and take the system to have a total of $L$ sites. As we will discuss
later, for our numerical calculations we will allow $\beta$ to deviate in 
a minimal way from $(\sqrt{5} - 1)/2$ in order to have periodic boundary
conditions. We will set $\hbar = 1$ in this paper.

It is possible to do a unitary transformation which changes $b_j \to - b_j$ 
but keeps $a_j$ unchanged for all $j$ in Eq.~\eqref{ham1}. This allows us 
to change the relative signs of both the time-independent and time-dependent 
hoppings on alternating bonds. Hence there are two possibilities for 
these two hoppings: both can be uniform, or one can be uniform and the
other staggered. We find it convenient to take the 
the driving term to be uniform rather than staggered. [This may also be
easier to realize experimentally. One way to drive the hopping is to
apply a time-dependent periodic pressure on the system. This would make 
all the bond lengths increase and decrease alternately with time. As
a result all the nearest-neighbor hoppings will change periodically
with time. This corresponds to a uniform term in the time-dependent
hopping]. Assuming the time-dependent hopping to be uniform, the 
time-independent hoppings ($\ga_1$ and $\ga_2$) can be either uniform
or staggered. In this paper, we will mainly study 
the staggered hopping model in which $\ga_2 = - \ga_1$. We will 
also briefly study the case of uniform hopping with $\ga_2 = \ga_1$
and compare the results found there with those obtained
for $\ga_2 = - \ga_1$. Finally, we note that if there is no
driving, i.e., $a=0$, then the cases of uniform and staggered time-independent
hoppings ($\ga_2 = \ga_1$ and $\ga_2 = - \ga_1$ respectively) are identical
to each other since they are related by the unitary transformation described
above; both cases reduce to the Aubry-Andr\'e model.

We now compute the Floquet Hamiltonian using first-order Floquet 
perturbation theory (FPT) for $a, ~\om \gg \ga_{1}, ~\ga_{2}, ~V_{0}$.~\cite{rev8,erg1,haldar,soori} FPT is a method for deriving 
the Floquet Hamiltonian in the limit where the driving amplitude is much
larger than the coefficients of the time-independent terms in the
Hamiltonian. In that limit, FPT is more powerful than the Magnus expansion 
which is often used in the high-frequency limit~\cite{rev4,rev6};
unlike the Magnus expansion, FPT does not involve expanding in 
powers of $1/\om$. To develop FPT for our model, 
we write the Hamiltonian as $H (t) = H_0 (t) + V_1 + V_2$, where, 
in terms of a momentum $k$ (which lies in the range $[-\pi/2, \pi/2]$), 
\bea H_{0} (t) ~=~ \sum_{k} ~[ a \sin (\om t) (1+e^{-2ik} ) 
a_{k}^{\dagger}b_{k} ~+~ {\rm H.c.}], \label{h0} \eea
and the perturbation has two parts,
\bea V_{1} &=& \sum_{k} ~[(\ga_{1}+\ga_{2}e^{-2ik}) a_{k}^{\dagger}
b_{k} + {\rm H.c.}], \non \\
V_{2} &=& \sum_{k,k'} ~[f_{1} (k,k') a_{k}^{\dagger}a_{k'} + f_{2} 
(k,k') b_{k}^{\dagger}b_{k'}], \label{v12} \eea
with
\bea f_{1} (k,k') &=& \frac{2 V_{0}}{L} \sum_{j} \cos [4\pi\beta 
(j-\frac{1}{2})]~ e^{-2i(k-k')j}, \non\\
f_{2} (k,k')&=& \frac{2 V_{0}}{L} \sum_{j} \cos [4\pi\beta j]~
e^{-2i(k-k')j}. \label{f12} \eea
We note that the Fourier transform of the quasiperiodic potential couples 
different momenta which is a consequence of the fact that such a potential 
breaks translation symmetry. More specifically, $f_1 (k,k')$ and 
$f_2 (k,k')$ are non-zero whenever $k-k' = \pm 2 \pi \beta$ mod $\pi$.

The instantaneous eigenvalues of $H_0 (t)$ are given by 
\bea E_{k \pm} &=& \pm ~2a ~\sin (\om t) \cos (k). \eea 
These satisfy the condition 
\bea e^{i \int_{0}^{T}dt [E_{k+}(t) - E_{k-} (t)]} =1, \eea
where $T = 2 \pi/\om$ is the time period of the drive.
We therefore have to carry out degenerate FPT.
The eigenfunctions corresponding to $E_{k\pm}$ are given by 
\bea \ket{k \pm} &=& \frac{1}{\sqrt{2}} \left(\begin{array}{cc}
1 \\ \pm e^{ik} \end{array}\right), \label{eigen1} \eea

We begin with the Schr\"odinger equation 
\beq i\frac{d\ket{\psi}}{dt} ~=~ \left( H_{0}+V \right)\ket{\psi},
\label{schr1} \eeq 
where $V=V_1 + V_2$. We assume that $\ket{\psi (t)}$ has the expansion
\bea \ket{\psi(t)} ~=~ \sum_{n} ~c_{n}(t)
~e^{-i\int_{0}^{t} dt' E_{n}(t')} \ket{n}. \label{psi1} \eea 
Eq.~\eqref{schr1} then implies that
\beq \frac{dc_{m}}{dt} ~=~ -i ~\sum_{n} ~\bra{m} V\ket{n} ~e^{i\int_{0}^{t} 
dt' ( E_{m}(t')-E_{n}(t'))} c_{n}. \label{cm1} \eeq
Integrating this equation, and keeping terms only to first order in
$V$, we find that 
\bea c_{m}(T) &=& c_{m}(0) ~- ~i ~\sum_{n}\int_{0}^{T}dt\bra{m} V\ket{n} 
\non \\
&& \times ~e^{i\int_{0}^{t} \left(E_{m}(t')-
E_{n}(t')\right)~dt'}c_{n}(0). \label{cm2} \eea 
This can be written as 
\bea c_{m}(T) ~=~ \sum_{n}\left(I-iH_{F}^{(1)}T\right)_{mn}c_{n}(0), 
\label{cm3} \eea 
where $I$ denotes the identity matrix and
$H_F^{(1)}$ is the Floquet Hamiltonian to first order in $V$. 

The calculation proceeds as follows. First, we consider $H_{F1}^{(1)} = 
\sum_k H_{Fk1}^{(1)}$ which is proportional to $V_{1}$. This term has 
non-zero matrix elements between states with the same momenta, and it yields
\bea \bra{k \pm} H_{Fk1}^{(1)} \ket{k \pm}&=& \pm \left(\ga_{1}+\ga_{2} \right)~
\cos (k),\non\\
\bra{k+}H_{Fk1}^{(1)} \ket{k-}&=&-i\left( \ga_{1}-\ga_{2} \right)
\sin (k) ~J_{0} \left(\mu_k\right) ~e^{i\mu_{k}},\non\\ 
\mu_k &=& \frac{4 a \cos (k)}{\om}. \label{matel}\eea
Using Eq.\ \eqref{matel}, we write $H_{Fk1}^{(1)}$ as
\bea H_{Fk1}^{(1)} &=& (\ga_{1}+\ga_{2}) \cos (k) ~\si_z \non \\
&& +~ (\ga_{1}-\ga_{2})
J_{0}\left(\mu_k\right) \cos(\mu_{k})\sin (k)~\si_y\non\\
&& +~ (\ga_{1}-\ga_{2})
J_{0}\left(\mu_k\right)\sin(\mu_{k})\sin (k)~\si_x. \eea
where we have defined the Pauli matrices $\si_{x,y,z}$ in the 
$\ket{k+},~\ket{k-}$ basis.

Next, we change basis to
\bea \ket{k \uparrow} &=&a_{k}^{\dagger}\ket{0}, \quad
\ket{k \downarrow} = b_{k}^{\dagger}\ket{0}, \eea so that \bea
\ket{k \pm} &=& \frac{1}{\sqrt{2}}\left( \ket{k\uparrow} \pm e^{ik}
\ket{k \downarrow} \right). \label{changebasis} \eea
Using a two-component operator $\psi_k = (a_k,b_k)$, we can write 
$H_{F1}^{(1)}$ in the new basis as 
\bea H_{F1}^{(1)} &=& \frac{1}{2} ~\sum_k \psi_k^{\dagger} 
\left( \begin{array}{cc} 
\al_{2k} & \al_{1k} \\ \al_{1k}^{\ast} & -\al_{2k} \end{array} \right) 
\psi_k \non\\
\al_{1k} &=& g_{1k} ~[1+J_0(\mu_k)~\cos(\mu_{k})] \non \\
&& +~ g_{2k} ~[1-J_0(\mu_k)~\cos(\mu_{k})],\non\\
g_{1k} &=& \ga_1 + \ga_2 e^{-2 i k}, \quad 
g_{2k} = \ga_2 + \ga_1 e^{-2 i k},\non\\
\al_{2k}& = &(\ga_{1}-\ga_{2})~\sin(k)~J_{0}(\mu_{k})~\sin(\mu_{k}).
\label{fhterm1} \eea 
We note that if $\ga_2 = \ga_1$, we obtain $\al_{1k} = 2 \ga_1 (1 + 
e^{-i2k})$ and $\al_{2k} = 0$. Hence $H_{F1}^{(1)}$ does not depend on 
the driving parameters $\om$ and $a$. It is possible that higher order 
terms will depend on the driving parameters when $\ga_2 = \ga_1$ but 
these terms will be smaller than the first-order term derived here. 
We therefore expect that driving will have a smaller effect when
$\ga_2 = \ga_1$ compared to $\ga_2 = - \ga_1$.

Next, we compute the first-order contribution to the Floquet Hamiltonian, 
$H_{F2}^{(1)}$, arising from $V_2$. To this end, we first define
\bea p_{\pm} (k,k') &=& \frac{1}{2} \left(f_{1}(k,k') \pm
f_{2}(k,k')e^{-i(k-k')}\right), \non \\
q_{\pm}(k,k') &=& \frac{1}{2} \left(f_{2}(k,k') \pm 
f_{1}(k,k')e^{i(k-k')}\right), \non \\
r_{\pm}(k,k') &=& \frac{i}{2} \left( f_{2}(k,k')e^{-ik}\pm 
f_{1}(k,k')~e^{-ik'}\right), \non \\
t_{\pm}(k,k') &=& \frac{i}{2} \left( f_{1}(k,k')e^{ik}\pm 
f_{2}(k,k')~e^{ik'}\right), \non \\
\mu_{kk'}^\pm &=& \frac{1}{2} ~(\mu_k ~\pm~ \mu_{k'}). \eea
We then find that in the $|k \pm \ra$ basis, the matrix elements of 
$H_{F2}^{(1)}$ are given by
\beq \la k s | H_{F2}^{(1)} | k' s' \ra ~=~ p_{ss'} (k,k')~
J_0 (\mu_{kk'}^{-ss'}) ~e^{i s \mu_{kk'}^{-ss'}}, \label{matel1} \eeq
where $ss' = +1$ if $s=s'=+1$ or $s=s'=-1$, and $ss'=-1$ if $s=+1, ~s'=-1$ or
$s=-1, ~s'=+1$.


Using Eq.~\eqref{changebasis}, we find that $H_{F2}$ is diagonal in the 
$|k \uparrow \ra$, $|k \downarrow \ra$ basis and takes the form 
\begin{widetext}
\bea H_{F2}^{(1)} &=& ~~\left[ \cos (\mu_{kk'}^-) ~J_{0}(\mu_{kk'}^-)~
p_{+}(k,k') ~+~ \cos (\mu_{kk'}^+) ~J_{0}(\mu_{kk'}^+) ~p_{-}(k,k') \right] ~
a_{k}^{\dagger}a_{k'} \non \\
&& + ~\left[ \cos (\mu_{kk'}^-) ~J_{0}(\mu_{kk'}^-)~ q_{+}(k,k') ~+~ \cos (\mu_{kk'}^+) ~J_{0}(\mu_{kk'}^+) ~q_{-}(k,k') \right] ~
b_{k}^{\dagger}b_{k'} \non \\
&& +~ \left[ \sin (\mu_{kk'}^-) ~J_{0}(\mu_{kk'}^-)~
r_{+}(k,k') ~+~ \sin (\mu_{kk'}^+) ~J_{0}(\mu_{kk'}^+) ~r_{-}(k,k') \right]~ a_{k}^{\dagger}b_{k'} \non \\
&& +~ \left[ \sin (\mu_{kk'}^-) ~J_{0}(\mu_{kk'}^-)~ t_{+}(k,k') ~+~ \sin
(\mu_{kk'}^+) ~J_{0}(\mu_{kk'}^+) ~t_{-}(k,k') \right] ~b_{k}^{\dagger}a_{k'}. 
\label{matel2} \eea
\end{widetext}
This completes our derivation of the first-order Floquet Hamiltonian within FPT.


We can follow the same procedure to find the Floquet Hamiltonian for
square pulse driving. We consider a driving protocol having the form
\bea f(t)~&=&~a~~~{\rm for}~~~0\leq t\leq T/2,\non\\
&=&-a~~~{\rm for}~~~T/2\leq t\leq T. \label{ft} \eea
The form of the first-order Floquet Hamiltonian for square pulse driving 
is similar to the form we obtained for sinusoidal driving except that
$J_{0}(x)$ has to be replaced by $\sin(x)/x$ and $\mu_{k}=4a\cos(k)/\om$ 
has to be replaced by $\mu_{k}=a T \cos(k)$. A square pulse driving is 
easier to study numerically since the Floquet operator $U$ which 
time-evolves the system for one time period $T$ can be obtained by simply
multiplying two operators, one which time-evolves from $t=0$ to $T/2$ and
the other which evolves from $t=T/2$ to $T$.

We now look at the form of $H_F^{(1)}$ in the limits $\om \gg a \gg \ga_1,
~\ga_2, ~V_0$ and $a \gg \om \gg \ga_1, ~\ga_2, V_0$, which we will refer 
to as the high- and intermediate-frequency regimes respectively. In both cases,
we are assuming that $\om \gg \ga_1, ~\ga_2, V_0$. If this condition is not
satisfied (i.e., if we are in the low-frequency regime), the Floquet 
Hamiltonian $H_F$ cannot be uniquely defined since 
the eigenvalues $e^{-i \ta_m}$ of the Floquet operator $U$ will not satisfy
$|\ta_m| \ll \pi$ for all the Floquet eigenstates $m$. Hence $H_F = (i/T) 
\ln U$ will suffer from branch cut ambiguities. This necessitates 
folding back of Floquet eigenstates into the first Floquet Brillouin 
zone~\cite{erg3}, a technical complication which we will avoid in this work. 

We first consider the case of sinusoidal driving and look at the two
limits separately.

\noi (i) The high-frequency limit can be studied using either the FPT
described above or the Magnus expansion. In the FPT, we have $\mu_k = 
(4 a /\om) \cos (k) \to 0$ for all $k$ lying in the range $[-\pi/2,\pi/2]$. 
Since the Bessel function satisfies $J_0 (z) \to 1$ when $z \to 0$, we find 
that $H_{F1}^{(1)} + H_{F2}^{(2)}$ approaches the time-independent
part, $V_1 + V_2$, of the Hamiltonian $H(t)$ given in Eq.~\eqref{v12}.
Hence, the high-frequency limit should give the same results as the
undriven system with $a=0$. However, we will see later that
although this is true for the Floquet eigenvalues, it does not seem to
hold for the Floquet eigenstates.

\noi (ii) The intermediate-frequency limit cannot be studied using
the Magnus expansion, and we have to use the FPT. We now use the fact that
$J_0 (z)$ goes to zero as $\sqrt{2/(\pi z)} \cos (z - \pi/4)$
when $|z| \to \infty$. Furthermore, $\mu_k \to \pm \infty$
except near the special point $k = \pi/2$.
Similarly, $\mu_{kk'}^{\pm} = (1/2) (\mu_k \pm \mu_{k'}) \to \pm \infty$
except near the point $(k,k') = (\pi/2,\pi/2)$ and the line
$k = \pm k'$. However, we saw after Eq.~\eqref{f12} that the functions 
$f_1 (k,k')$ and $f_2 (k,k')$ are non-zero only if $k-k'= \pm 2 \pi \beta$
mod $\pi$. Since $\beta = (\sqrt{5}-1)/2 \simeq 0.618$, and $k, ~k'$
lie in the range $[-\pi/2,\pi/2]$, we take $2 \pi \beta$ mod $\pi$ to be 
equal to either $(\sqrt{5}-2) \pi \simeq 0.742$ or $(\sqrt{5}-3) \pi \simeq 
- 2.400$, so that $k-k' \simeq 0.742$ or $-2.400$. Putting together the
conditions arising from $\mu_{kk'}^{\pm}$, $f_1 (k,k')$ and $f_2 (k,k')$, 
we see that in Eq.~\eqref{matel2}, the terms involving products of $J_0 
(\mu_{kk'}^\pm)$ and $f_1 (k,k')$ or $f_2 (k,k')$ will go to zero at all
values of $k, ~k'$ except near the four points $(k,k') \simeq \pm (0.742/2) 
(1,-1)$ or $\pm (2.400/2) (1,-1)$. Ignoring these special points, we see 
that in the intermediate-frequency limit, $H_{F2}^{(1)} \to 0$ in 
Eq.~\eqref{matel2}, which implies that the effect of the quasiperiodic
potential goes to zero. Similarly, we see that except near the special 
point $k=\pi/2$, $J_0 (\mu_k) \to 0$, and $H_{F1}^{(1)}$ 
in Eq.~\eqref{fhterm1} approaches the form
\beq H_{F1}^{(1)} ~=~ \frac{1}{2} ~(\ga_1 + \ga_2) ~\sum_k 
\psi_k^{\dagger} \left( \begin{array}{cc} 
0 & 1 + e^{-i2k} \\ 1 + e^{i2k} & 0 \end{array} \right) \psi_k.
\label{fhterm3} \eeq
This describes a system with uniform hopping amplitude $(1/2) (\ga_1 + 
\ga_2)$ on all bonds. We thus see that in the intermediate-frequency limit, 
the system approaches a simple limit in which the hopping amplitude is 
uniform and there is no quasiperiodic potential. The quasienergies will then
be given by $(\ga_1 + \ga_2) \cos (k)$ which is a gapless spectrum.

We reach similar conclusions in the case of square pulse driving. 
Namely, the high-frequency limit is the same as an undriven 
system with staggered hopping amplitudes $\ga_1, ~\ga_2$ and a 
quasiperiodic potential $V_0$, while the intermediate-frequency limit is a 
system with a uniform hopping amplitude $(1/2) (\ga_1 + \ga_2)$
and no quasiperiodic potential.

Finally, we discuss some exact symmetries of the Floquet Hamiltonian. 
We can show that the Floquet Hamiltonian for the present problem can 
only have terms proportional to odd powers of $V_{1}$ and $V_{2}$ 
for a drive satisfying the condition $f(-t)=-f(t)$ 
(which is satisfied if $f(t)$ is either given by $a\sin(\om t)$ or by the
square pulse form in Eq.~\ref{ft}). To see this, we first show that the 
Floquet operator $U$ satisfies
\bea U^{-1}(a,\ga_{1},\ga_{2},V_{0})~=~U(a,-\ga_{1},-\ga_{2},-V_{0}). 
\label{sym1} \eea
Hence the Floquet Hamiltonian $H_F$, which satisfies $U = e^{- i H_F T}$,
must satisfy
\bea H_{F}(a,\ga_{1},\ga_{2},V_{0})=-H_{F}(a,-\ga_{1},
-\ga_{2},-V_{0}). \label{sym2} \eea
Hence $H_{F}$ can only have odd powers of $\ga_1, ~\ga_2$ and $V_0$.
As a result, the next correction after the terms of first-order in 
$\ga_1, ~\ga_2$ and $V_0$ will be of third order and will therefore be
small. Hence, in the following sections we will only show the 
perturbative results based on first-order FPT.

\section{Results for the model with $~\ga_2 = - \ga_1$}
\label{sec3}

The Floquet Hamiltonian obtained in Sec.~\ref{sec2} allows us to study the
properties of the Floquet eigenstates within first-order FPT. To this end, 
we numerically diagonalize $H_F^{(1)} = H_{F1}^{(1)} + H_{F2}^{1)}$
to obtain its eigenvectors and eigenvalues 
\bea H^{(1)}_F |\psi_m^{(1)}\ra &=& E_{Fm}^{(1)} |\psi_m^{(1)}\ra . \eea
On the other hand, the Floquet operator $U$ will have eigenvectors 
$| \psi_m \ra$ and corresponding eigenvalues $\exp (-i \ta_m)$, where 
$\ta_m = E_{Fm} T$ can be chosen to lie in the range $[-\pi,\pi]$.
Then the quasienergies $E_{Fm}$ obtained from $\ta_m$ will lie
in the range $[-\om/2, \om /2]$.

\begin{figure}[htb]
\subfigure[]{\includegraphics[width=0.5\linewidth]{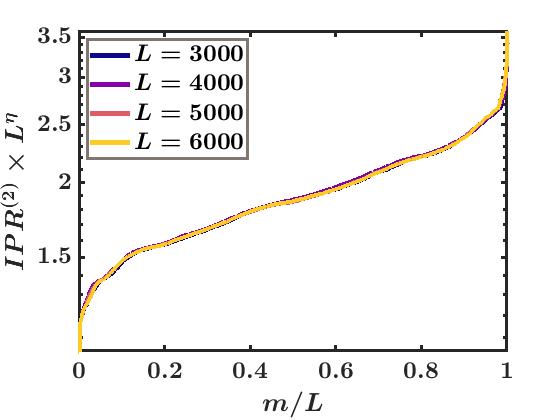}}%
\subfigure[]{\includegraphics[width=0.5\linewidth]{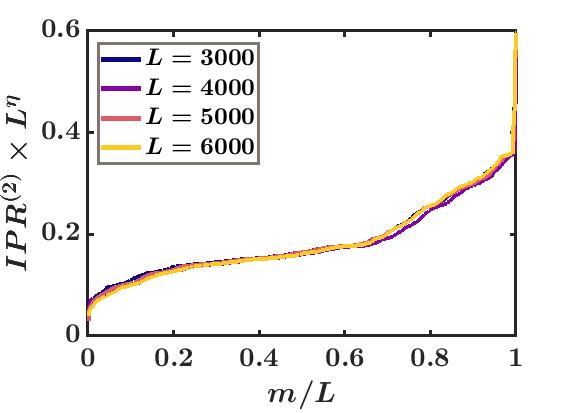}}\\%
\subfigure[]{\includegraphics[width=0.5\linewidth]{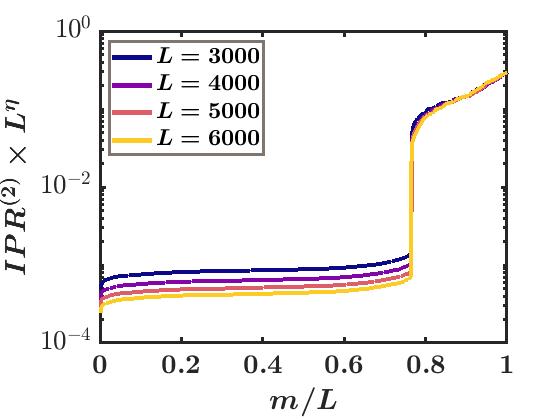}}%
\subfigure[]{\includegraphics[width=0.5\linewidth]{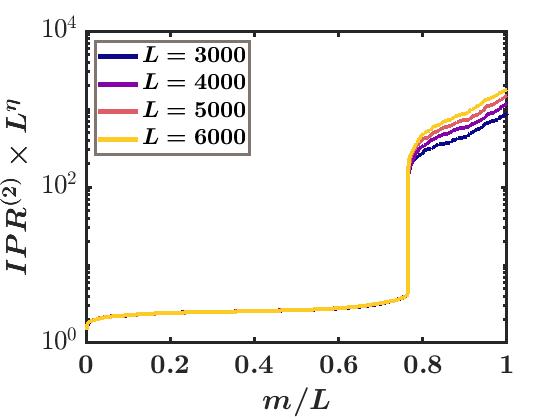}}\\%
\subfigure[]{\includegraphics[width=0.55\hsize]{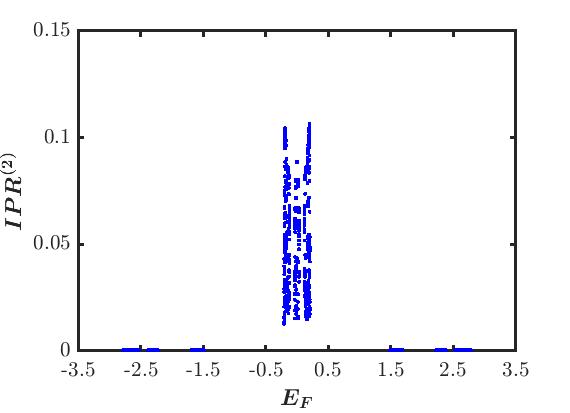}}%
\caption{(a-d) Plots of $I_m^{(2)} L^\eta$ (sorted in increasing order)
versus $m/L$ for $\ga_1=1$,
$\ga_2 =-1$, $V_0=2.5$, system sizes $L = 3000, ~4000, ~5000$ and 6000,
and sinusoidal driving. (a) $I_m^{(2)} L^{\eta}$ for Floquet 
eigenstates with $\om=5$ and $a=5$. The exponent $\eta = 1$, and we see 
that all states are extended. (b) $I_m^{(2)} L^{\eta}$ with $\om=40$ 
and $a=5$. Here $\eta = 0$ showing that all states are localized. 
(c) and (d) $I_m^{(2)} L^{\eta}$ with $\om=15.1$ and $a=5$. Plots (c)
and (d) correspond to $\eta=0$ and 1 respectively, and they clearly 
show a jump in the IPR value around $m/L \simeq 0.76$. (e) Plot of $I_m^{(2)}$
versus quasienergy $E_F$ for $\om = 15$, $a=5$, and $L=3000$. We see several 
gaps in $E_F$.}
\label{fig01} \end{figure}

\begin{figure}[htbp]
\subfigure[]{\includegraphics[width=0.5\linewidth]{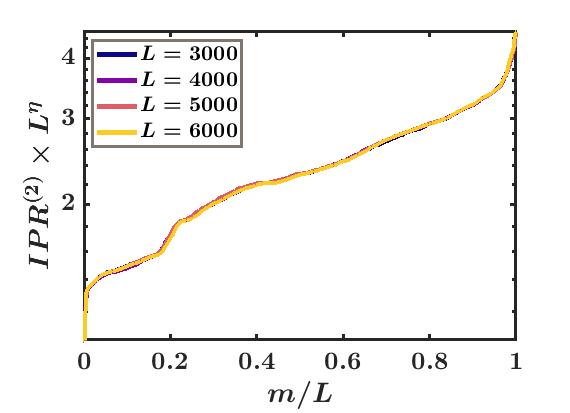}}%
\subfigure[]{\includegraphics[width=0.5\linewidth]{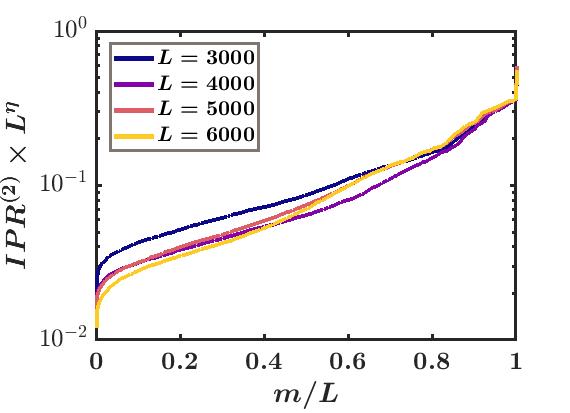}}\\
\subfigure[]{\includegraphics[width=0.5\linewidth]{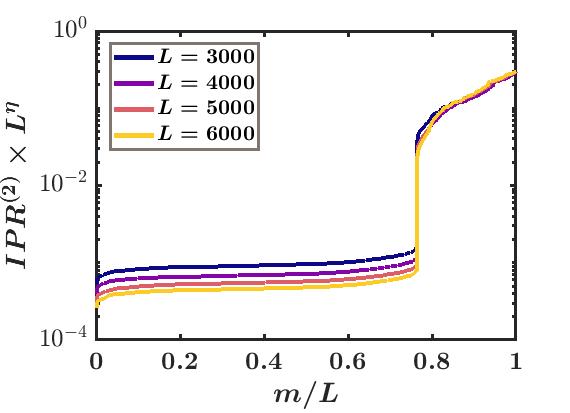}}%
\subfigure[]{\includegraphics[width=0.5\linewidth]{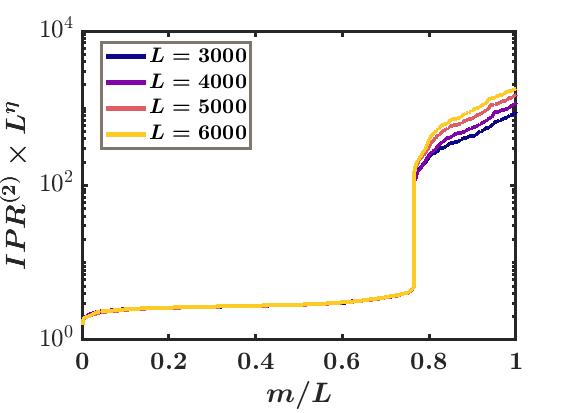}}\\
\subfigure[]{\includegraphics[width=0.55\linewidth]{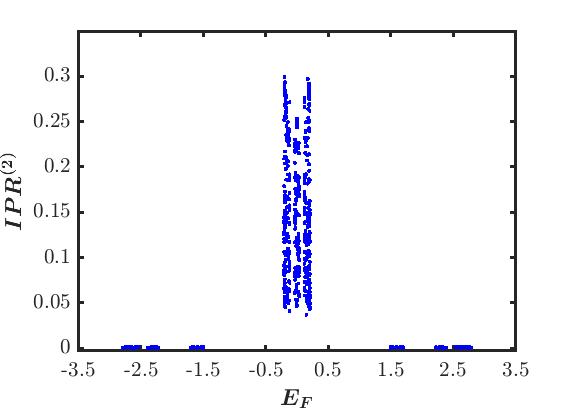}}%
\caption{(a-d) Plots of $I_m^{(2)} L^\eta$ versus $m/L$ and (e) plot
of $I_m^{(2)}$ versus $E_F$ for the same 
parameter values as in Fig.\ \ref{fig01} but obtained using first-order FPT. 
All the plots agree qualitatively with the plots from exact 
numerics shown in Fig.~\ref{fig01}.} \label{fig02} \end{figure}

The results obtained using FPT will be compared with those obtained from 
an exact numerical calculation of the Floquet operator. To this end, 
we divide the time period $T$ into $N$ steps so that $H(t)$ does not 
vary appreciably within the time step $\de t= T/N$. For the present
case, we find that $N \sim 500$ achieves this task; a further increase in 
$N$ does not change the numerical results appreciably. This allows 
us to use the standard Suzuki-Trotter decomposition to write $U$ as 
\bea U(T,0) &=& \prod_{j=1,N} U(t_j,t_{j-1}), \non\\
U_j \equiv U(t_j,t_{j-1}) &=& e^{- i H[(t_j+t_{j-1})/2] \de t}, 
\label{trodec} \eea
where $t_0=0$ and $t_N = T$. We can then numerically diagonalize $U$ to 
obtain its eigenvectors $| \psi_m \ra$ and eigenvalues 
$e^{-i \ta_m}$. We can then write
\beq U(T,0) ~=~ \sum_m e^{-i \ta_m} | \psi_m \ra \la \psi_m |. 
\label{uresol} \eeq 
In this section, we will study and compare the properties of the Floquet
eigenstates obtained using first-order FPT and exact numerics. To this
end, we compute the inverse participation ratio (IPR) of these 
eigenstates given, for a normalized Floquet eigenstate $m$, as
\beq I_m^{(2)} ~=~ \sum_{j=1}^L |\psi_m (j)|^4.
\label{iprex} \eeq
It is well-known that $I_m^{(2)}$ can distinguish between localized and 
extended states; $I_m^{(2)} \sim L^{-\eta}$ where $\eta=0 ~(1)$ for 
localized (extended) eigenstates. 

\subsection{Localized, extended and multifractal states}
\label{sec3a}

We now present our numerical results. Figures~\ref{fig01} (a-d) 
show plots of $I_m^{(2)} L^{\eta}$ (sorted in increasing order)
versus $m/L$ for systems with $\ga_1 = 1$, $\ga_2 = -1$,
$V_0 = 2.5$, different values of $\om$ and $a$, and sinusoidal driving. 
For generating this and all subsequent figures, the quasiperiodic
potential at site $j$ is taken to be $V_j = V_0 \cos (2 \pi \al 
j)$, where $\al$ is taken to be a rational approximant for $(\sqrt{5}-1)/2$ 
by choosing $\al = N/L$, where $L$ is the system size, and $N$ is the 
integer closest to $L (\sqrt{5}-1)/2$. We have chosen 
$L = 3000, ~4000, ~5000$ and 6000.
The plots clearly indicate a jump in the IPR value for certain 
ranges of values of $\om$ and $a$. From Figs.\ \ref{fig01} (a) and (b), 
we find that at relatively low (high) driving frequencies, 
all the Floquet eigenstates are extended (localized). This can be seen from the
collapse of the $I_m^{(2)} L^{\eta}$ curves for $\eta=1$ at $\om = 5$ 
(Fig.\ \ref{fig01} (a)) and $\eta=0$ at $\om = 40$ (Fig.\ \ref{fig01} (b)) 
for different $L$. In contrast, at an intermediate driving frequency, 
$\om =15.1$, as shown in Figs.\ \ref{fig01} (c) and (d), a jump in the IPR
between localized and extended eigenstates around $m/L \simeq 0.76$. The data 
for different $L$ collapses upon scaling with $L^{\eta}$ for two different
values of $\eta$; the collapse happens for $\eta=0$ for $m/L \gtrsim 
0.764$ 
(Fig.\ \ref{fig01} (c)) and for $\eta=1$ for $m/L \lesssim 0.764$ 
(Fig.\ \ref{fig01} (d)). This demonstrates the appearance of an IPR
jump around $m/L \simeq 0.76$ at this driving frequency. Figure~\ref{fig01} (e)
shows a plot of $I_m^{(2)}$ versus the quasienergy $E_F$ for $\om = 5$,
$a=5$, and $L=3000$. This clearly shows that there are gaps in the quasienergy
spectrum; we call these mobility gaps. (In this paper, we use the term 
mobility gap regardless of whether the states in the quasienergy bands on the 
two sides of a gap are both extended or both localized or one extended and 
the other localized). The states with $E_F$ around
zero have large IPRs and correspond to localized states, whereas states
with $E_F$ away from zero (on either the positive or the negative side) 
have very small IPRs and are extended states. [It is important to note that 
sorting in increasing order of $I_m^{(2)}$ (as done in Figs.~\ref{fig01} (a-d))
and in increasing order of $E_F$ (as done in Fig.~\ref{fig01} (e)) are
quite different from each other. The presence of mobility gaps
becomes apparent only in the latter way of sorting].

In Fig.~\ref{fig02}, similar plots are shown for Floquet eigenstates obtained 
from first-order FPT. These correspond to the same parameter values 
as their counterparts in Fig.\ \ref{fig01}. A
comparison between Figs.\ \ref{fig01} and \ref{fig02} shows that 
first-order FPT provides a reasonable match to the exact
numerics, although the IPR values found from exact numerics are
significantly smaller than the ones found from first-order FPT. Importantly, 
both exact numerics (Fig.~\ref{fig01} (e)) and first-order FPT 
(Fig.~\ref{fig02} (e)) show mobility 
gaps at intermediate driving frequencies like $\om = 5$. We conclude
that first-order FPT can provide a good understanding of the driven
system when $a \gg \gamma_1, \gamma_2$.

\begin{figure}[!tbp]
\subfigure[]{\includegraphics[width=0.5\hsize]{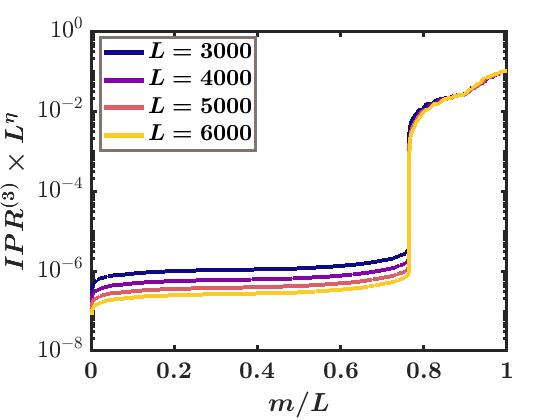}}%
\subfigure[]{\includegraphics[width=0.5\hsize]{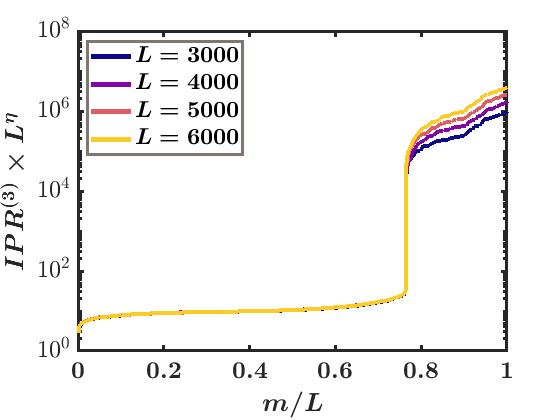}}\\
\subfigure[]{\includegraphics[width=0.5\hsize]{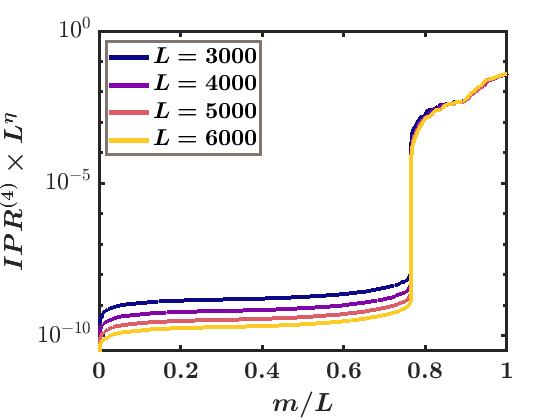}}%
\subfigure[]{\includegraphics[width=0.5\hsize]{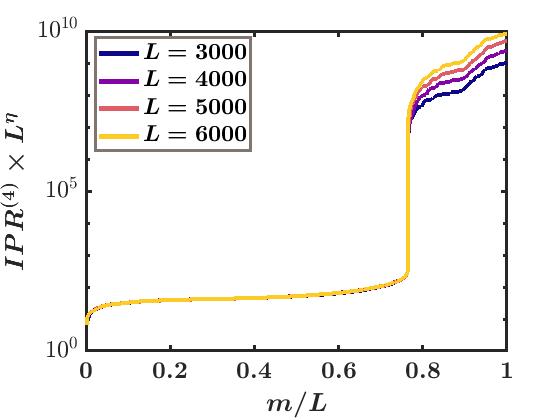}}\\
\caption{Plots showing $I_m^{(3)} L^\eta$ and $I_m^{(4)} L^\eta$ 
(sorted in increasing order) versus $m/L$ obtained by exact 
numerical calculations for $L=3000, ~4000, ~5000$ and 6000, with 
$\ga_1 = 1, ~\ga_2 = -1, ~a=5, ~\om =15.1$, $V_0 = 2.5$, and sinusoidal 
driving. Plot (a) shows a data collapse of $I_m^{(3)}$ with
$\eta = 0$ for $m/L \gtrsim 0.764$. Plot (b) shows a data collapse of 
$I_m^{(3)}$ with $\eta = 2$ for $m/L \lesssim 0.764$. Plot (c) 
shows a data collapse of $I_m^{(4)}$ with $\eta = 0$ for $m/L
\gtrsim 0.764$. Plot (d) shows a data collapse of $I_m^{(4)}$ 
with $\eta = 3$ for $m/L \lesssim 0.764$. These plots show that 
$I_m^{(q)}$ scales with $\eta = 0$ for states with $m/L \gtrsim
0.764$ (localized states) and with $\eta = q -1$ for states with $m/L 
\lesssim 0.764$ (extended states).} \label{fig03} \end{figure}

To determine if any of the eigenstates exhibit multifractal behavior,
we calculate a generalized IPR defined as
\beq I_m^{(q)} ~=~ \sum_j ~|\psi_m (j)|^{2q}, \label{imp} \eeq
and study its scaling with the system size 
$L$.~\cite{sarkar1,caste,evers,rodri} 
If $I_m^{(q)}$ scales as $L^{-\eta_q}$ and $\eta_q = (q-1) D_q$, 
then $D_q = 1$ for all $q$ for extended states (since $|\psi_m (j)|^2$ 
is of the order of $1/L$ for all $j$ for such states), and 
$D_q = 0$ for all $q$ for localized states (since $|\psi_m (j)|^2$
for such states is of order 1 over a finite region whose size
remains constant as $L \to \infty$). Multifractal states 
typically have $0 < D_q < 1$ for all $q$. Figure~\ref{fig03} shows 
plots of $I_m^{(q)} L^\eta$ for $q=3$ and 4 for systems with 
$\ga_1 =1$, $\ga_2 = -1$, $\om = 15.1$, $a=5$, $V_0 = 2.5$, different system 
sizes, and sinusoidal driving. Figures~\ref{fig03} (a) and (c) show that
states with $m/L \gtrsim 0.764$ scale with $L$ with the powers $\eta_3 =
\eta_4 = 0$, showing that $D_q = 0$; hence these states are localized.
Figures~\ref{fig03} (b) and (d) show that
states with $m/L \lesssim 0.764$ scale with powers $\eta_3 = 2$ and
$\eta_4 = 3$, showing that $D_q = 1$; hence these states are extended.
Thus there is no evidence of multifractal states for this set of
parameter values. However, we will show below that multifractal behavior
appears for a different range of values of $\om$.

\begin{figure}[!tbp]
\subfigure[]{\includegraphics[width=0.5\hsize]{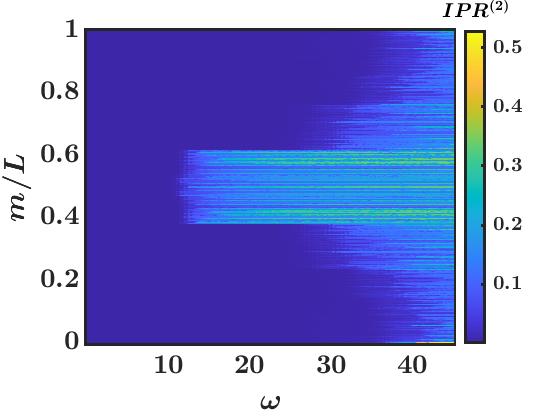}}%
\subfigure[]{\includegraphics[width=0.5\hsize]{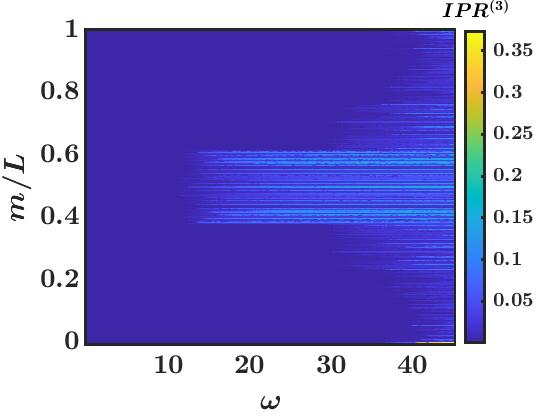}}\\
\subfigure[]{\includegraphics[width=0.5\hsize]{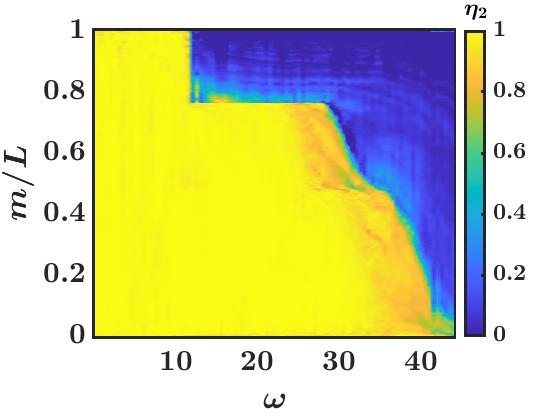}}%
\subfigure[]{\includegraphics[width=0.5\hsize]{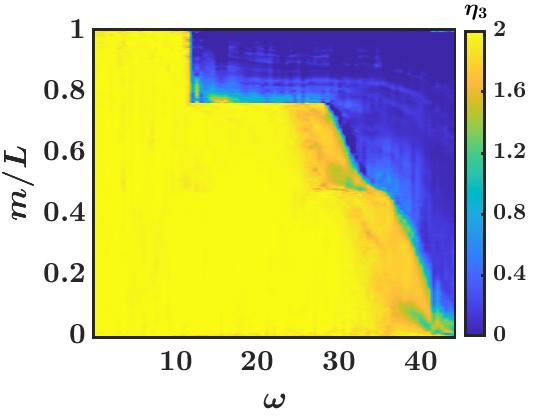}}\\
\subfigure[]{\includegraphics[width=0.5\hsize]{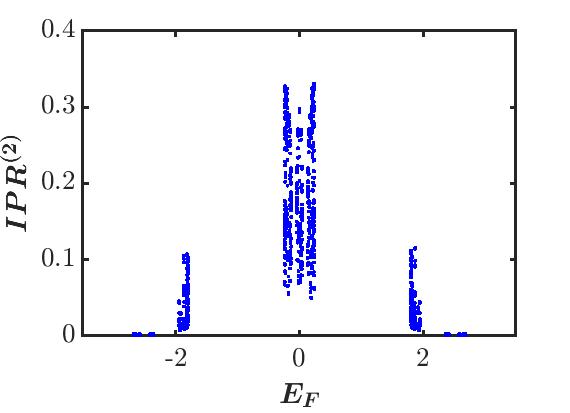}}
\caption{(a-b) Plots of $I_m^{(2)}$ and $I_m^{(3)}$ (sorted 
in increasing order of quasienergy $E_{Fm}$) versus $\om$ and $m/L$ 
for $\ga_{1}=1$, $\ga_{2}=-1$, $a=5$, $V_{0}=2.5$, $L=3000$, and 
square pulse driving. (c-d) Plots of $\eta_{2}$ and $\eta_{3}$ 
(extracted from the IPRs sorted in increasing order for 
$L=3000, ~4000, ~5000$ and 6000) versus $\om$ and $m/L$, 
with the parameter values same as in plots (a-b). Multiple jumps in 
$I_m^{(2)}$ and $I_m^{(3)}$ observed in plots (a-b) 
indicate multiple localization-delocalization transitions. This is 
more evident in plots (c-d) where $\eta_{2}$ and $\eta_{3}$ are 
extracted from the scaling analysis of IPR values. 
We note that for $30 \lesssim \om \lesssim 45$ in plots 
(c-d), the values of $\eta_{2}$ and $\eta_{3}$ for a small
fraction of states are different from the usual scaling 
exponents of both extended states ($\eta_2=1$ and $\eta_{3}=2$)
and localized states ($\eta_2 = \eta_3 = 0$).
Plot (e) shows $I_m^{(2)}$ as a function of $E_{Fm}$ (sorted
in increasing order) for $\ga_{1}=1$, $\ga_{2}=-1$, $a=5$, $V_{0}=2.5$, $\om=30$ and $L=3000$.} 
\label{fig05} \end{figure}

In Fig.~\ref{fig05}, we show plots of $I_m^{(2)}$ and $I_m^{(3)}$ 
versus $m/L$, and their scaling exponents with $L$, $\eta_2$ and $\eta_3$,
versus $\om$ and $m/L$ for systems with $\ga_1 = 1$, $\ga_2 = -1$, 
$a=5$, $V_0 = 2.5$, and square pulse driving.
Figures~\ref{fig05} (a-b) show $I_m^{(2)}$ and $I_m^{(3)}$
for a system with $L=3000$, while Figs.~\ref{fig05} (c-d) show the
scaling exponents $\eta_2$ and $\eta_3$ for these two quantities
extracted from the results for $L=3000, ~4000, ~5000$ and 6000.
We see that most states have $\eta_2 = \eta_3 = 0$ (localized)
at lower frequencies and have $\eta_2 = 1$ and $\eta_3 = 2$ (extended)
at higher frequencies. Interestingly, however, we see some states
for which $\eta_2$ and $\eta_3$ are different from the values of
both localized and extended states. Hence these states have
a multifractal nature. In Fig.~\ref{fig05} (e) we show $I_m^{(2)}$
versus the quasienergy $E_{Fm}$ for $\om = 30$. For this value of $\om$,
all three kinds of states coexist, extended, multifractal and localized,
as is evident from the scaling exponents $\eta_2$ and $\eta_3$ shown in 
Figs.~\ref{fig05} (c-d).
We have also studied what happens if the driving is sinusoidal instead 
of square pulse and we find similar results which we do not show here.
This demonstrates that the generation of multifractal 
states does not depend significantly on the driving protocol.


\begin{figure}[!tbp]
\subfigure[]{\includegraphics[width=0.5\hsize]{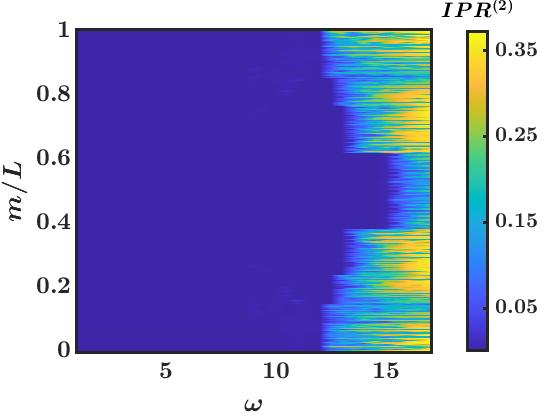}}%
\subfigure[]{\includegraphics[width=0.5\hsize]{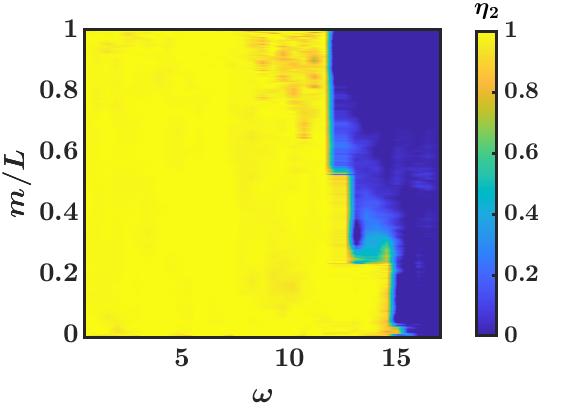}}\\
\caption{(a) Plot of $I_m^{(2)}$ (sorted in increasing order of quasienergy)
versus $\om$ and $m/L$ with $\ga_{1}=\ga_{2}=1$ (uniform hopping), 
$a=5$, $V_{0}=2.5$, $L=3000$, and square pulse driving. (b) Plot of 
$\eta_{2}$ (extracted from the IPRs sorted in increasing order 
for $L=3000, ~4000, ~5000$ and 6000) versus $\om$ and $m/L$.}
\label{fig06} \end{figure}

Figure~\ref{fig06} shows plots of $I_m^{(2)}$ and the scaling exponent
$\eta_2$ 
versus $\om$ and $m/L$ for $L = 3000, ~4000, ~5000$ and 6000, for 
$\ga_1 = \ga_2 =1$ (uniform hopping), $a=5$, $V_0 = 2.5$, and square 
pulse driving. The results looks somewhat similar to those shown in
Fig.~\ref{fig05} for $\ga_1 = - \ga_2 = 1$ (staggered hopping) in that 
both extended and
localized states appear in the two cases. However, extended states 
persist up to a larger values of $\om$ for $\ga_1 = - \ga_2 = 1$
compared to $\ga_1 = \ga_2 = 1$.

It is useful to look at the average values of both $I_m^{(2)}$ and the
normalized participation ratio (NPR) which, up to a factor of $L$, is the
inverse of $I_m^{(2)}$.~\cite{li} For the $m$-th Floquet eigenstate, the 
NPR is defined as ${\rm NPR}_m = 1/(L I_m^{(2)})$. The average values of 
these two quantities are then defined as $\la {\rm IPR} \ra = (1/L)
\sum_m I_m^{(2)}$ and $\la {\rm NPR} \ra = (1/L) \sum_m {\rm NPR}_m$. 
For a large system size $L$, we have $I_m^{(2)} \sim 1/L ~(1)$ and 
${\rm NPR}_m \sim 1 ~(1/L)$ for an extended (localized) state $m$,
respectively. Hence, the quantity $\phi = \log_{10} (\la {\rm IPR} \ra 
\la {\rm NPR} \ra)$ will be large and negative
(of the order of $- \log_{10} L$) either if all states are extended or if all
states are localized. But if $\phi$ is not large and negative, this would 
indicate that there are some states which are neither extended nor localized. 
We will also look at the average value of the Shannon entropy as another 
measure of the degree of localization~\cite{dynloc4}. For the $m$-th 
Floquet eigenstate, the Shannon entropy is defined as $S_{m}=-\sum_{n}
|\psi_{m} (n)|^{2} \ln (|\psi_{m} (n)|^2)$, and its average is then given 
by $\la S \ra =(1/L) \sum_{m}S_{m}$). For an extended (localized) 
state $m$, we have $S_m \sim \ln L ~(0)$, respectively. Hence the average 
value $\la S \ra$ will be of order $\ln L$ if all states are 
extended and will be close to zero if all states are localized.

\begin{figure}[!tbp]
\subfigure[]{\includegraphics[width=0.5\hsize]{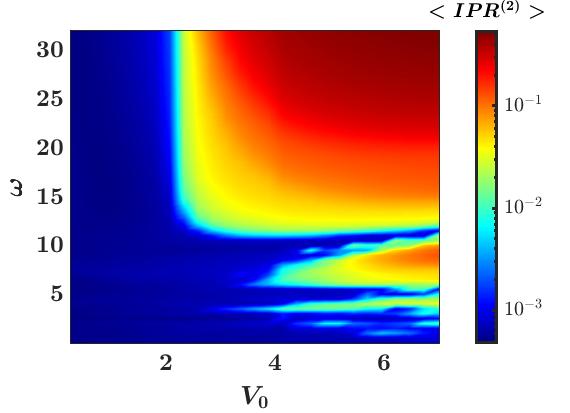}}%
\subfigure[]{\includegraphics[width=0.5\hsize]{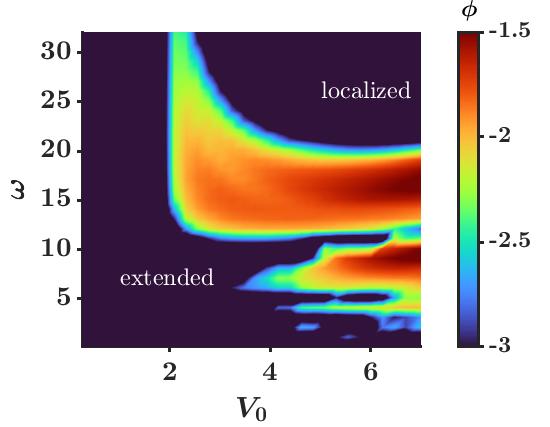}}\\
\subfigure[]{\includegraphics[width=0.5\hsize]{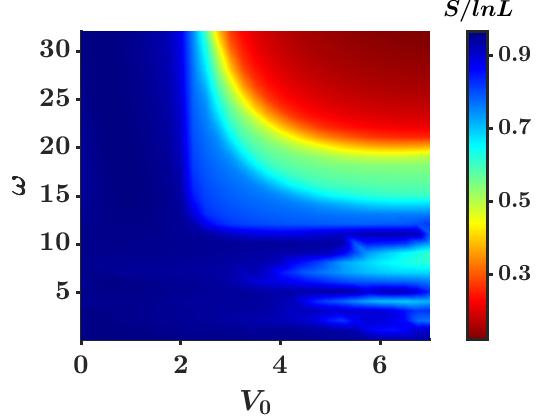}}%
\caption{(a) Plot on a log scale of the average of $I_m^{(2)}$,
denoted as $\la IPR \ra$, versus $V_0$ and $\om$ for $\ga_{1}=1,
\ga_{2}=-1$, $a=5$, $L=3000$, and square pulse driving. 
(b) Plot of $\phi = \log_{10} ({\la \text{IPR} \ra \la
\text{NPR} \ra})$ versus $V_0$ and $\om$. (c) Plot of
the average Shannon entropy $\la S \ra$ 
versus $V_0$ and $\om$. Plots (a-c) suggest that all the Floquet 
eigenstates below $V_{0} \simeq 2$ are extended for all values of $\om$. 
For $2 \lesssim V_{0} \lesssim 3.5$, we find three distinct regions, 
with only extended states for small $\om$, coexisting states of different
types for intermediate $\om$, and only localized states for large $\om$. 
A different behavior appears for $V_{0}\gtrsim 3.5$, 
where we see multiple re-entrant transitions between regions with 
only extended and only localized states as $\om$ is increased.}
\label{fig07} \end{figure}

Figure~\ref{fig07} shows plots of $\la {\rm IPR} \ra$, $\phi =
\log_{10} (\la {\rm IPR} \ra \la {\rm IPR} \ra$), and
$\la S \ra$ versus $\om$ and $V_0$, for a system with $\ga_1 = 
1$, $\ga_2 =-1$, $a=5$ and square pulse driving. Figure~\ref{fig07} (a)
indicates that all states are extended when both $\om$ and
$V_0$ are small and are localized when both $\om$ and $V_0$ are large. 
For $2 \lesssim V_0 \lesssim 3.5$, we see that only extended states 
low values of $\om$, coexisting states of different types
for intermediate values of $\om$,
and only localized states for large values of $\om$. Interestingly, when 
$V_0 \gtrsim 3.5$, we find multiple re-entrant transitions between regions 
with only extended and only localized states as $\om$ is increased.
These observations are confirmed qualitatively by Figs.~\ref{fig07} (b) 
and (c). We note that the range of values of $V_0$ which supports 
re-entrant transitions is the same range where the time-independent part
of the model exhibits only localized states.

\begin{figure}[!tbp]
\subfigure[]{\includegraphics[width=0.5\hsize]{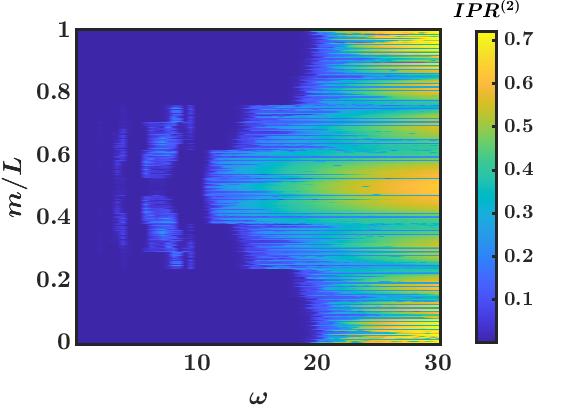}}%
\subfigure[]{\includegraphics[width=0.5\hsize]{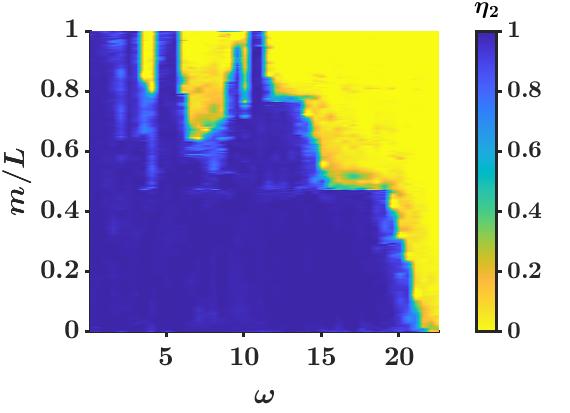}}
\caption{(a) Plot of $I_m^{(2)}$ (sorted in increasing order of
quasienergy) versus $\om$ and $m/L$, with $\ga_{1}=1$, $\ga_{2}=-1$, $a=5$,
$V_{0}=5.5$, $L=3000$, and square pulse driving. 
(b) Plot of $\eta_2$ versus $\om$ and $m/L$. 
We extract $\eta_{2}$ using system sizes $L=3000, ~4000, ~5000$ and 6000. 
The plots confirm the multiple transitions seen in Fig.~\ref{fig07} for
$V_0 \gtrsim 3.5$ as $\om$ is varied. Plot (b) shows several transitions.
A transition to a phase with all extended states occurs at $\om \simeq
5.1$, and this phase then continues up to $\om \simeq 7.1$. For $\om 
\gtrsim 7.1$, we get both extended and localized states till $\om \simeq
9.1$. Another phase of extended states appears for $9.2 \lesssim \om 
\lesssim 12.1$. Beyond $\om \simeq 12.1$, plot (b) suggests that the 
system moves towards complete localization as $\om$ increases.}
\label{fig08} \end{figure}

Figure~\ref{fig08} shows plots of $I_m^{(2)}$, the scaling exponent 
$\eta_2$ and the real part of the Floquet eigenvalues versus $\om$ 
and $m/L$ for systems with $\ga_1 = 1$, $\ga_2 = -1$, $a=5$, $V_0 = 
5.5$ and different sizes, for square pulse driving. This figure 
confirms the multiple re-entrant transitions between regions which 
are completely extended, completely localized, or have both extended 
and localized states, that we see in Fig.~\ref{fig07} when 
$V_0 \gtrsim 3.5$. (We note that similar re-entrant transitions have
been observed in a kicked quasicrystal~\cite{grover}). We would like
to mention here that the plots in Fig.~\ref{fig08} are difficult to 
explain by any perturbation theory since the values of $a$, $\om$ and
$V_0$ are all comparable to each other.

\begin{figure}[!tbp]
\subfigure[]{\includegraphics[width=0.5\hsize]{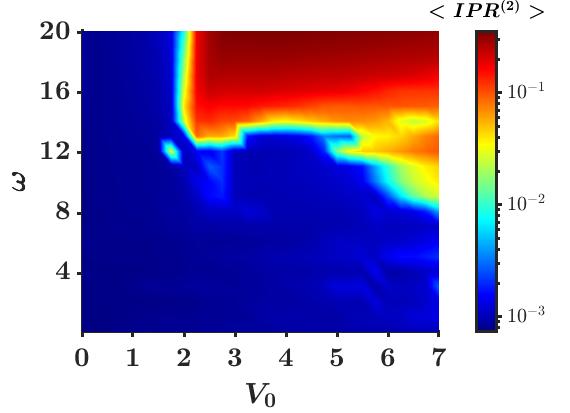}}%
\subfigure[]{\includegraphics[width=0.5\hsize]{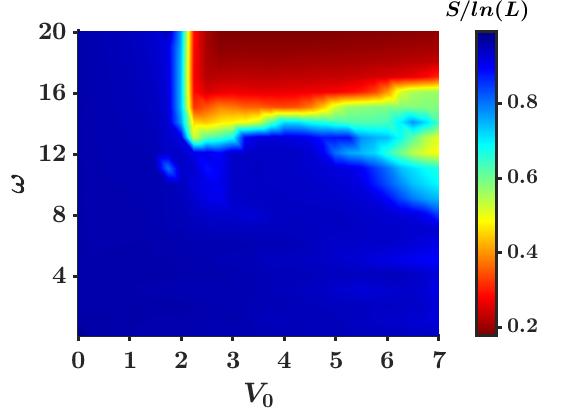}}\\
\caption{(a) Plot on a log scale of the average of $I_m^{(2)}$, 
$\la {\rm IPR} \ra$, versus $\om$ and $V_{0}$, with
$\ga_{1}=\ga_{2}=1$ (uniform hopping), $a=5$, $L=3000$, and square 
pulse driving. (b) Plot of average Shannon entropy, $\la S \ra$,
versus $\om$ and $V_{0}$. For $V_{0} \gtrsim 3.5$, we see that the 
case of uniform hopping shown here deviates significantly 
from the case of non-uniform hopping shown in Fig.~\ref{fig07}.}
\label{fig10} \end{figure}

Fig.~\ref{fig10} shows the average $\la {\rm IPR} \ra$ and 
average Shannon entropy $\la S \ra$ as functions of $V_0$ and
of $\om$ for a system with uniform hopping, $\ga_1 = \ga_2 = 1$, $a=5$
and $L=3000$, for square pulse driving. The results deviate 
significantly from the case of staggered hopping, $\ga_2 = - \ga_1$, shown
in Fig.~\ref{fig07}, particularly for $V_0 \gtrsim 3.5$.

It is known that there are no re-entrant transitions in the Aubry-Andr\'e
model (which is the time-independent part of our model) if there is no driving.
Furthermore, there are no such transitions in a driven
model with no quasiperiodic potential. Comparing Fig.~\ref{fig07} 
(where $\ga_2 = - \ga_1$) and Fig.~\ref{fig10}
(where $\ga_2 = \ga_1$), we see that staggered hopping, driving and 
quasiperiodic potential are all necessary to clearly see re-entrant transitions
between delocalized states and other kinds of states.

\subsection{High-frequency limit}
\label{sec3b}

In the high-frequency limit $\om \gg a \gg \ga_1, ~\ga_2, ~V_0$, we 
expect the effects of driving to disappear as we have argued in Sec.~\ref{sec2}
Surprisingly, however, we find that
although the quasienergies for large $\om$ match the energies
for the undriven system, the IPRs do not agree in the two cases.

\begin{figure}[!tbp]
\subfigure[]{\includegraphics[width=0.5\hsize]{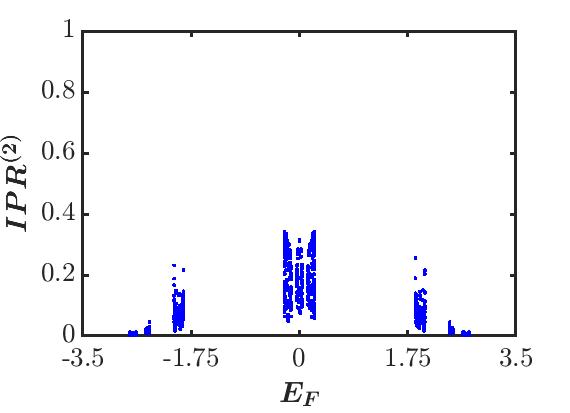}}%
\subfigure[]{\includegraphics[width=0.5\hsize]{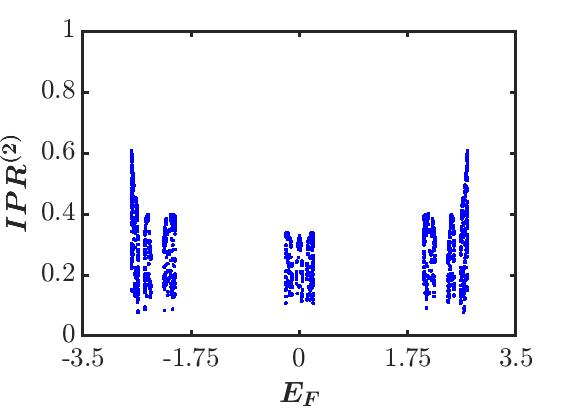}}\\
\subfigure[]{\includegraphics[width=0.5\hsize]{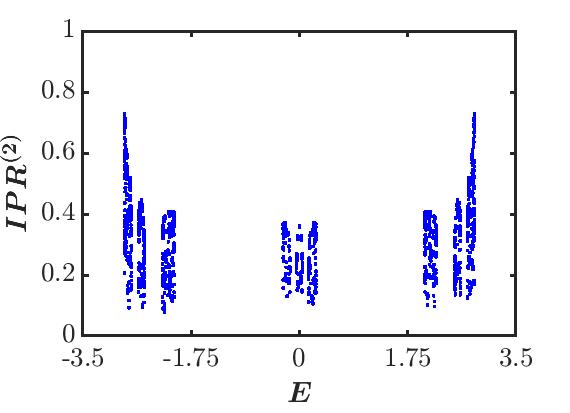}}%
\subfigure[]{\includegraphics[width=0.5\hsize]{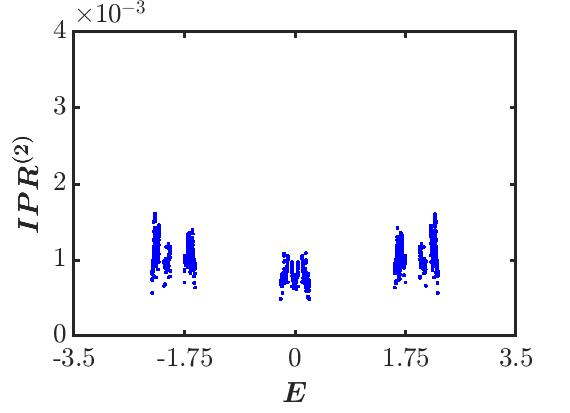}}
\caption{(a) Plot of $I_m^{(2)}$ versus quasienergy $E_F$ (sorted in 
increasing order) for a system with $\ga_1 = 1$, $\ga_2 = -1$, 
$\om =35$, $a=5$, $V_0 = 2.5$, $L=3000$, and square pulse driving.
(b) Plot of $I_m^{(2)}$ versus quasienergy for a system with 
$\ga_1 = 1$, $\ga_2 = 1$ (uniform hopping), $\om =35$, $a=5$, and
$V_0 = 2.5$. (c) Plot of $I_m^{(2)}$ versus energy $E$ (sorted in
increasing order) for an undriven system ($a=0$) with $\ga_1 = 1$, 
$\ga_2 = 1$, and $V_0 = 2.5$. (d) Plot of $I_m^{(2)}$ versus energy $E$ 
(sorted in increasing order) for an undriven system ($a=0$) with 
$\ga_1 = 1$, $\ga_2 = -1$, and $V_0 = 1.5$. Note that the range of IPR
values in plot (d) is much smaller than in plots (a-c). In all cases 
we have considered system size $L=3000$ and square pulse driving.}
\label{fig11} \end{figure}

In Figs.~\ref{fig11} (a) and (b), we show plots of $I_m^{(2)}$ versus
quasienergy $E_F$ (sorted in increasing order) for systems with $\ga_1 = 1$,
$\ga_2 = -1$ and $+1$ respectively, $\om =35$, $a=5$, $V_0 = 2.5$, 
$L=3000$, and square pulse driving. Fig.~\ref{fig11} (c) shows a plot
of $I_m^{(2)}$ versus energy $E$ (sorted in increasing order) for an
undriven system with $\ga_1 = 1$, $\ga_2 = 1$ (the relative sign of
$\ga_1$ and $\ga_2$ is unimportant in the absence of driving), 
$V_0 = 2.5$, and $L=3000$. We have chosen $V_0 = 2.5$ so that
$V_0 > 2 \ga_1$ and all the states of the undriven system are 
localized. We note that the quasienergies in plots (a) and (b) 
and the energies in plot (c) have almost the same values.
This is expected since driving at high frequencies should give the
same Floquet Hamiltonian as the undriven part of the Hamiltonian.
Surprisingly, however, the IPR values are not the same in the three
plots. The IPR values for the driven system with uniform hopping amplitudes
differ only a little from the IPRs of the undriven system, but
the IPRs of the driven staggered model ($\ga_2 = - \ga_1$) differ
substantially from those of the undriven system. (This may be 
because driving has a smaller effect when $\ga_2 = \ga_1$ compared
to $\ga_2 = - \ga_1$, for reasons explained after Eq.~\eqref{fhterm1}).
We observe that the undriven system (plot (c)) shows only localized 
states, but the driven staggered model (plot (a)) shows that some of the 
states have very low IPR values and are therefore extended states.
Thus driving even at a high frequency seems to convert some of the 
localized states to extended states, provided that $a \gg \ga_1, ~\ga_2$ 
and $V_0$. (We have checked numerically that if $a \ll \ga_1, ~\ga_2$
and $V_0$, then all the states remain localized). This suggests 
that although terms of order $1/\om$ and higher in the Floquet 
Hamiltonian are small for large $\om$, they have a significant
effect on the IPR values of the Floquet eigenstates. It is possible
that these small terms have a negligible effect on the quasienergies,
but they couple localized states which lie close to each other and
this hybridization produces extended states.

\begin{figure}[htb]
\subfigure[]{\includegraphics[width=0.5\hsize]{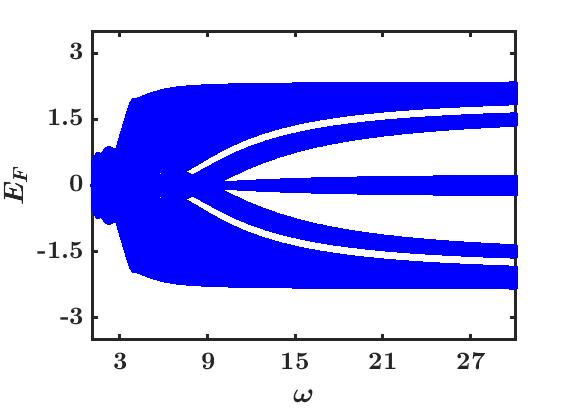}}%
\subfigure[]{\includegraphics[width=0.5\hsize]{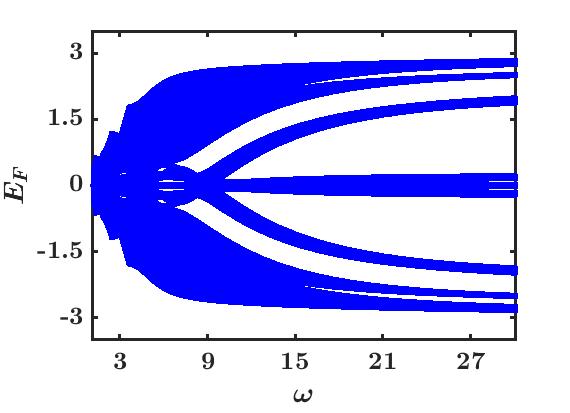}}\\
\subfigure[]{\includegraphics[width=0.5\hsize]{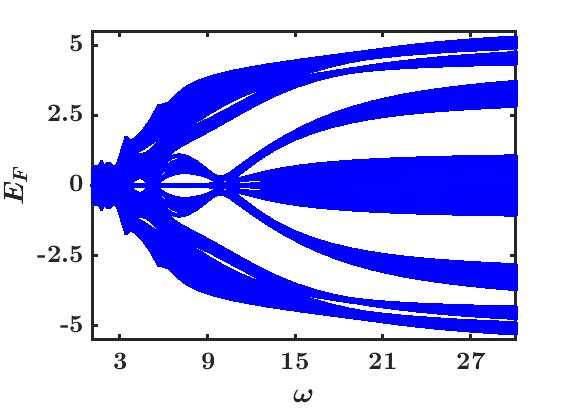}}
\caption{Plots of $E_{Fm}$ versus for $\om$ for (a) $V_0 = 1.5$,
(b) $V_0 = 2.5$, and (c) $V_0 = 5.5$, for systems with $\ga_1 = 1$, 
$\ga_2 = -1$, $a=5$, size $L=2000$, and square pulse driving. 
The spectrum of $E_{Fm}$ is gapless for intermediate $\om$ and develops 
several gaps as $\om$ increases.} \label{fig12} \end{figure}

On the other hand, if we choose $V_0 < 2 \ga_1$ and all the other parameters 
are the same as in Figs.~\ref{fig11} (a-c), we find that there is no
discernible differences between the spectra of quasienergies (or
energies) and IPR values in the three cases. All the states
are extended in all three cases (driven with $\ga_2 = \pm \ga_1$ and
undriven). Fig.~\ref{fig11} (d) shows a plot of the energies and 
corresponding IPR values for an undriven system with $\ga_1 = 1$, 
$\ga_2 = -1$, and $V_0 = 1.5$. The IPR versus quasienergy plots for 
the driven systems with $\ga_2 = \pm \ga_1$ look the same and are not 
shown here.

In Fig.~\ref{fig12}, we show plots of all the quasienergies $E_{Fm}$ for 
systems with $\ga_1 = 1$, $\ga_2 = -1$, $a=5$, size $L=2000$, square 
pulse driving, and (a) $V_0 = 1.5$, (b) $V_0 = 2.5$, and (c) $V_0 = 5.5$.
We have chosen these three values of $V_0$ for the following reason. 
For $V_0 = 1.5$, all states are found to be extended. For $V_0 = 2.5$,
various kinds of states can coexist depending on the values of $\om$,
as we see in Fig.~\ref{fig05}. For $V-0 = 5.5$, we find re-entrant transitions
as shown in Fig.~\ref{fig08}.
The plots in Fig.~\ref{fig12} show that the spectrum of $E_{Fm}$ is
generally gapless for intermediate frequencies but several gaps appear as 
$\om$ is increased. This can be qualitatively understood from the FPT as 
follows. In the limit that $\ga_1, ~\ga_2, ~V_0 \ll \om \ll a$, we saw in the
discussion given towards the end of Sec.~\ref{sec2} that the model reduces to
one in which the hopping amplitude is uniform and the quasiperiodic potential
is absent. This system clearly has no gaps in the spectrum. On the other hand,
when $\ga_1, ~\ga_2, ~V_0 \ll a \ll \om$, we saw that the model reduces to
an undriven system ($a=0$). Hence the quasienergies will approach the energies
of the undriven system. For the undriven Aubry-Andr\'e model, it is known that
there is a large number (in fact, an infinite number) of energy gaps
for any non-zero value of $V_0$.~\cite{aubry,soko} It would be useful 
to understand precisely how the different gaps appear successively as $\om$ 
is increased from intermediate to large values. [We have checked numerically
that for $V_0 = 0$, there are no gaps for any value of $\om$ or $a$. 
Hence the gaps seen in Fig.~\ref{fig12} are entirely due to the 
presence of a quasiperiodic potential].

\subsection{Floquet Hamiltonian in real space}
\label{sec3c}

\begin{widetext}

\begin{figure}[htb]
\subfigure[]{\includegraphics[width=0.32\hsize]{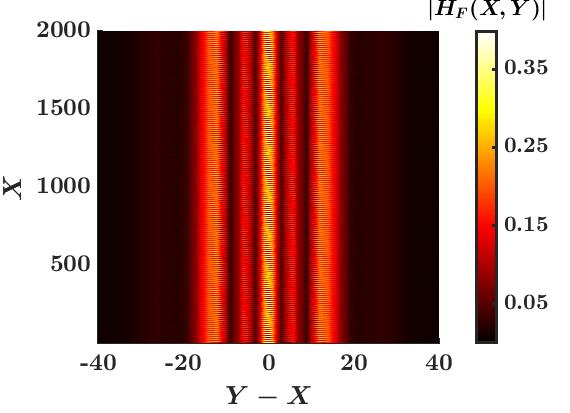}}%
\subfigure[]{\includegraphics[width=0.32\hsize]{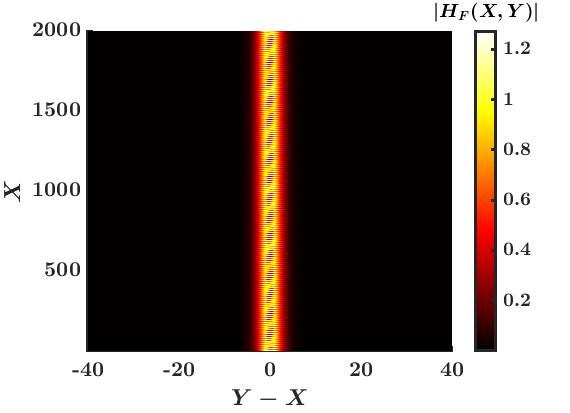}}%
\subfigure[]{\includegraphics[width=0.32\hsize]{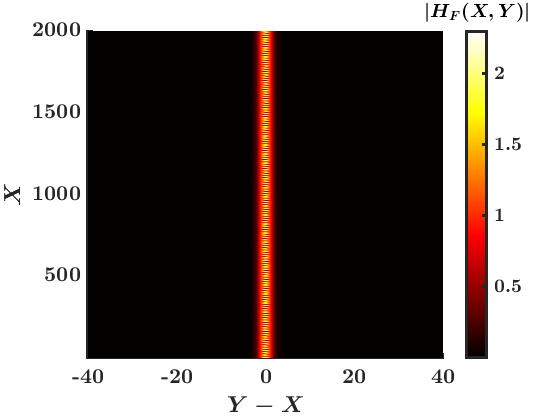}}\\
\caption{Plots of the absolute values of
the matrix elements of the Floquet Hamiltonian $H_{F} (X,Y)$ in 
real space ($X$ and $Y$ represent space coordinates), where the axes are
chosen to be $X$ and $Y-X$. In all cases, we considered $\ga_1 = 1,~ 
\ga_2 = -1$, $V_0 = 2.5$, system size $L = 2000$, and sinusoidal 
driving. Figures (a-c) are obtained from exact numerical calculations. 
The parameters used are (a) $a=10, ~\om = 5$, (b) $a=5, ~\om = 15.1$, 
and (c) $a=5, ~\om = 40$. 
We note that plot (a) and (d) is 
for a system with only extended states, plot (b) is
for a system with both extended and localized states,
and plot (c) is for a system with only localized states.} 
\label{fig13} \end{figure}

\end{widetext}

To visualize the effects of localization, it is instructive to look at the 
matrix elements of the Floquet Hamiltonian. Figure~\ref{fig13} shows the 
absolute values of the matrix elements of $H_F (X,Y)$ in real space for 
systems with $\ga_1 = 1$, $\ga_2 = - 1$, $V_0 =2.5$, size 
$L=2000$, and sinusoidal driving. $H_F (X,Y)$ is obtained by doing
a Fourier transform of $H_F$ in momentum space as found numerically
from the Floquet operator $U$ (Figs.~\ref{fig13} (a-c)).
The plots are shown as a function of $Y-X$ and $X$. If the system was 
translation invariant, the plot would depend only on the relative 
coordinate $Y-X$ and not on $X$. Indeed we see that the plots do not 
vary with $X$ at length scales much larger than $1/\beta \simeq 1.62$ 
(which is the length scale of variation of the quasiperiodic potential 
$\cos ( 2 \pi \beta j)$). In terms of the relative coordinate $Y-X$,
the plots show a large spread in Fig.~\ref{fig13} (a) corresponding 
to all states being extended, a smaller spread in Fig.~\ref{fig03} (b) corresponding to a case where there are both localized and extended
states, and a very small spread in Fig.~\ref{fig13} (c) 
corresponding to all states being extended. We find that $H_F (X,Y)$ 
obtained by doing a Fourier transform of $H_F$ in momentum space as 
found from the first-order FPT gives very similar results, and we do not
show those plots here.

A more detailed explanation of the results shown in Fig.~\ref{fig13} is as
follows. We know that although the Hamiltonian of the undriven Aubry-Andr\'e 
model is local in real space, it exhibits a delocalized phase below a critical disorder strength. We now look at the structure of matrix elements of the Floquet Hamiltonian $H_F (X,Y)$ in 
real space. In all the plots in Fig.~\ref{fig13}, 
we have taken we consider $\gamma_{1}=1$, $\gamma_{2}=-1$, $a=5$, and $V_{0}=2.5$. $V_{0}=2.5$ necessarily implies that the static model exhibits a localized phase in the absence of driving. However, in the 
presence of driving and with diminishing values of $\omega$, 
we observe a significant decrease in the value of the diagonal elements
$H_F (X,X)$. The values of these elements is found to be 
about 2 for $\omega$=40, about $1.2$ for $\omega$=15.1, and about 
$0.35$ for $\omega$=5. The decreasing
value of the diagonal elements implies that the effective strength of the
quasiperiodic potential is decreasing. Furthermore, we find that there is
a rapid spreading of the off-diagonal matrix elements signifying a
growing non-locality of $H_F$ as $\omega$ decreases. The combination of 
these two effects leads the system towards increasing delocalization.

\subsection{Spreading of a one-particle wave packet}
\label{sec3d}

\begin{figure}[!tbp]
\subfigure[]{\includegraphics[width=0.5\hsize]{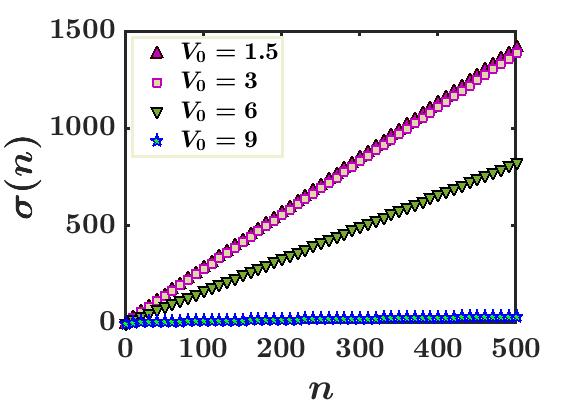}}%
\subfigure[]{\includegraphics[width=0.5\hsize]{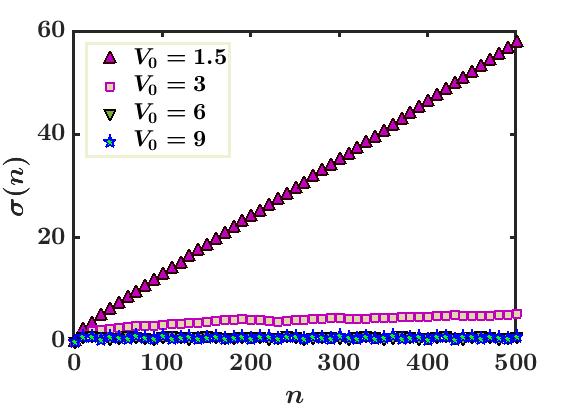}}\\
\subfigure[]{\includegraphics[width=0.5\hsize]{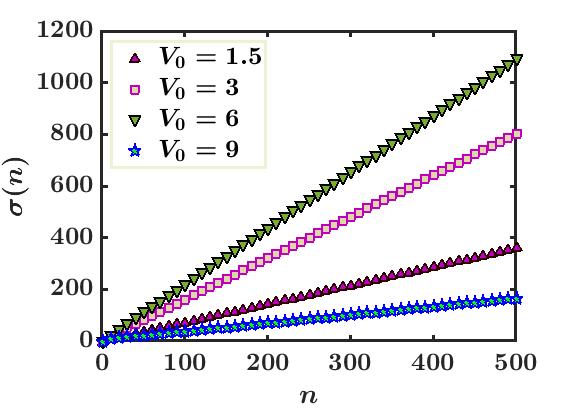}}%
\subfigure[]{\includegraphics[width=0.5\hsize]{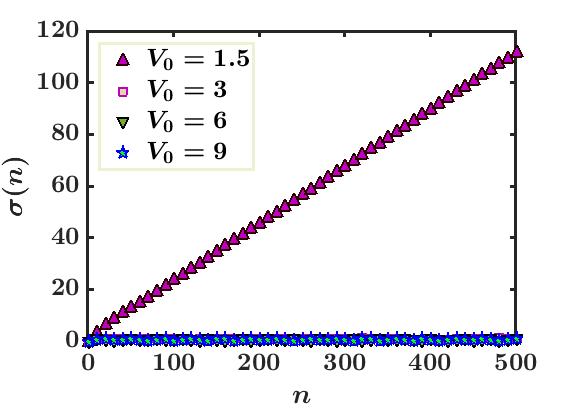}}\\
\caption{(a-b) Root mean squared displacement, $\si(n)$, versus the
stroboscopic evolution number $n$ for $\ga_{1}=1$, $\ga_{2}=-1$, $a=5$, and 
$\om=5$ and $\om=30$ respectively, for a one-particle state which is initially
located at the middle site of a large system. (c-d) Plot of $\si(n)$ 
versus $n$ for $\ga_{1}=\ga_{2}=1$ (uniform hopping),
$a=5$, and $\om=5$ and $\om=30$, respectively. All the figures are for system 
size $L=6000$ and square pulse driving. 
Plots (a-b) show that compared to the high-frequency regime (where the 
behavior is similar to an undriven system), the intermediate-frequency 
regime shows a ballistic behavior up to a larger value of 
$V_{0}$ for the case of staggered hopping compared to uniform hopping. For 
the uniform hopping system at the lower value of $\om=5$, plot (c) exhibits a 
non-monotonic ballistic velocity of $\sigma (n)$ as a function of $V_0$ before
showing a completely localized behavior for large $V_0$. Such a non-monotonic
behavior is not seen at the higher value of $\om=30$ as shown in plot (d).}
\label{fig14} \end{figure}

We now consider a measure of delocalization given by the spreading of 
a one-particles state which is initially localized at the middle site of 
system of size $L$, namely, the site $j_0 = L/2$. This initial state,
denoted $| \psi_{in} \ra$, is evolved over $n$ time periods by 
acting on it with the Floquet operator $n$ times. This gives 
\beq |\psi_{nT} \ra ~=~ U^n ~| \psi_{in} \ra. \eeq
The root mean squared displacement is then given by
\beq \si (n) ~=~ [\sum_j ~(j - j_0)^2 ~|\psi_{nT} (j)|^2]^{1/2}. \eeq
The growth of $\si (n)$ with $n$ gives an idea of how delocalized the 
system is. Fig.~\ref{fig14} shows plots of $\si (n)$ for systems with
$\ga_1 =1$, $\ga_2 = \pm 1$ (uniform and staggered hopping amplitudes
respectively), $a=5$, $\om = 5$ and 30, various values of $V_0$, 
and square pulse driving. (The system size $L$ is taken to be 6000).
We see that the spreading of the wave packet is always ballistic. The 
ballistic velocity, given by the slope of $\si (n)$ versus $n$, is smaller 
for $\om = 30$ compared to $\om = 5$. This is because driving 
with a very large frequency is equivalent to not driving at all.
In the absence of driving, it is known that a system with a quasiperiodic 
potential with strength $V_0 < 2 \ga_1$ has only extended states, while
a system with $V_0 > 2 \ga_1$ has only localized states~\cite{aubry,soko}; 
these results hold regardless of whether $\ga_2 = \ga_1$ or
$\ga_2 = - \ga_1$ since we can change the sign of $\ga_2$ by doing a unitary
transformation which does not affect $\ga_1$ and $V_0$. Indeed we see
in Figs.~\ref{fig14} (b) and (d), that the wave packet spreads ballistically 
if $V_0 < 2$ but remains completely localized if $V_0 > 2$. The situation
at the lower frequency, $\om =5$, is more interesting. The system with 
staggered hopping, $\ga_2 = - \ga_1$, shows ballistic spreading for $V_0 =
1.5, ~3$ and 6, and no spreading for $V_0 = 9$. However, the system with
uniform hopping, $\ga_2 = \ga_1$, always shows some ballistic spreading
for all values of $V_0$. Furthermore, the ballistic velocity is a non-monotonic
function of $V_0$, being smaller for $V_0 = 1.5$ and 9 compared to 
$V_0 = 3$ and 6. A similar non-monotonic behavior of the ballistic 
spreading is seen in Fig.~\ref{fig15} for driving at an intermediate
frequency with $\om = 12$. 

\section{Perturbation theory for a model with uniform hopping}
\label{sec4}

In this section we will study a system with a uniform hopping amplitude, 
$\ga_{1}=\ga_{2}=1$, which is driven by a square pulse. For uniform
hopping, the unit cell consists of a single site and we will denote
the fermionic operator as $c_j$.
The Hamiltonian is given by
\bea H(t)&=& (1+f(t)) \sum_{j} ~(c_{j}^{\dagger}c_{j+1}+
{\rm H.c.}) \non\\
&&+ ~V_{0} ~\sum_j \cos(2\pi\beta j) ~c_{j}^{\dagger}c_{j}, \eea
where $f(t)$ has the square pulse form described in Eq.~\eqref{ft}. 
Note that for $V_0 = 0$, the periodic driving would have no effect 
since the time-independent and driven parts of the Hamiltonian commute
with each other in that case, and the Floquet Hamiltonian $H_F$ would
just be given by $\sum_j (c_j^\dagger c_{j+1} + {\rm H.c.})$ independently
of the driving frequency. 

\begin{figure}[!tbp]
\subfigure[]{\includegraphics[width=0.5\hsize]{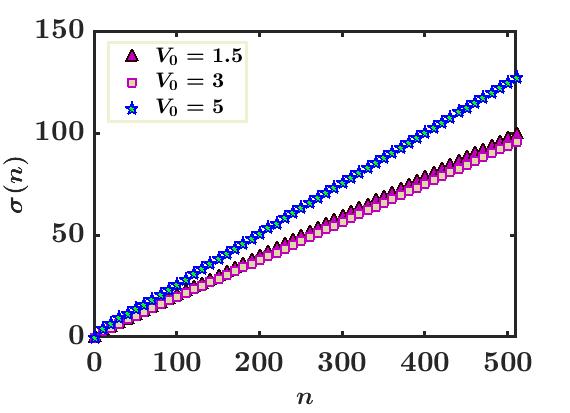}}%
\subfigure[]{\includegraphics[width=0.5\hsize]{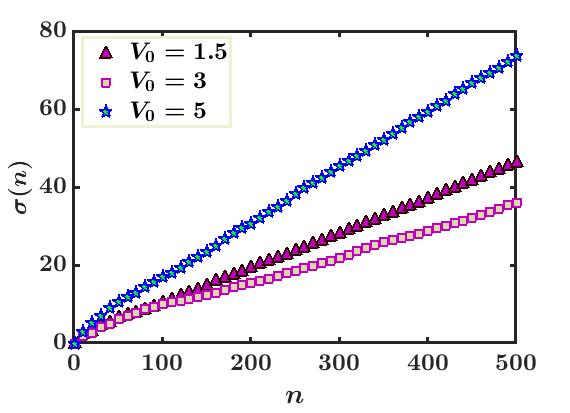}}\\
\caption{(a) Root mean squared displacement, $\si(n)$, versus the
stroboscopic evolution number $n$ from the exact numerical calculation 
with $\ga_{1}=\ga_{2}=1$ (uniform hopping), $a=5$, and $\om=12$. (b) $\si (n)$ 
versus $n$ as obtained from the second-order Floquet Hamiltonian in
Sec.~\ref{sec4}. Both figures are for $L=3000$ and square pulse driving. 
The two figures agree qualitatively and indicate a non-monotonic ballistic
velocity of $\si (n)$ as $V_{0}$ increases.} \label{fig15} \end{figure}

We will now find the Floquet Hamiltonian $H_F$ in the high-frequency regime 
using van Vleck perturbation theory~\cite{rev4,rev6}. This is a 
perturbative expansion in powers of $1/\om$, and it has the advantage over 
the Floquet-Magnus expansion that it does not depend on the phase of the
driving protocol, i.e., it is invariant under $f(t) \to f(t+t_0)$.
Furthermore, unlike the FPT which gives us $H_F$ in momentum space, the
van Vleck perturbation theory gives us $H_F$ in real space. This makes
it more suitable for studying the dynamics of a wave packet.

The zeroth-order Floquet Hamiltonian given by van Vleck perturbation theory
turns out to be
\bea H_{F}^{(0)} &=& \frac{1}{T} ~\int_0^T dt ~H(t) \non \\
&=& \sum_{j} (c_{j}^{\dagger}c_{j+1}+ {\rm H.c.}) 
+ V_{0} \sum_j \cos(2\pi\beta j) ~c_{j}^{\dagger}c_{j}. \non \\
&& \eea
The $m$-th Fourier component of $H(t)$ is given by
\bea H_{m}&=&\frac{2 i a}{m\pi} \sum_j (c_{j}^{\dagger}c_{j+1}+ {\rm H.c.})
~~ {\rm for} ~~ m ~~\text{odd},\non\\
&=& 0~~ {\rm for} ~~ m ~~\text{even ~and}~ \ne 0, \eea
and $H_0 = H_F^{(0)}$.
We then find that the first-order Floquet Hamiltonian, $H_F^{(1)} = \sum_{m 
\neq 0}\left[H_{-m},H_{m}\right]/(2m\om)$, is zero since $H_{-m} = (-1)^m H_m$
for all $m$. The second-order Floquet Hamiltonian 
consists of two terms
\bea H_{F}^{(2)}&=&\sum_{m\neq0}\frac{[[H_{-m},H_{0}],H_{m}]}{2m^{2}\om^{2}}
\non\\
&&+\sum_{m\neq0}\sum_{n\neq0}\frac{[[H_{-m},H_{m-n}],H_{n}]}{3mn\om^{2}}. 
\label{hf2} \eea
The first term in Eq.~\eqref{hf2} takes the form
\bea H_{F}^{(2)}&=&2t_{2}\sum_{j} ~[\cos(2\pi\beta(j+1))+\cos(2\pi\beta(j-1))
\non\\
&& ~~~~~~~~~~~~~~ -2\cos(2\pi\beta j)] ~c_j^\dagger c_j \non\\
&&+t_{2}\sum_{j} ~[2\cos(2\pi\beta(j+1))-\cos(2\pi\beta(j+2))\non\\
&& ~~~~~~~~~~~~~~ -\cos(2\pi\beta j)] ~(c_{j}^{\dagger}c_{j+2}+ 
{\rm H.c.}), \eea
where $t_{2}=a^{2}T^{2}V_{0}/96$. The second term in Eq.~\eqref{hf2} turns out 
to be zero. The third-order Floquet Hamiltonian vanishes due to the symmetry
given in Eq.~\eqref{sym2}. In principal we can calculate higher order terms 
but such corrections are expected to be small compared to $H_{F}^{(0)}$ and
$H_{F}^{(2)}$. 

Fig.~\ref{fig15} shows a comparison between the results obtained
for the root mean squared displacement $\si(n)$ versus the stroboscopic 
evolution number $n$ as obtained from exact numerics and the second-order 
Floquet Hamiltonian derived above, for systems with $a=5$,
$\om =12$, various values of $V_0$, and square pulse driving. 
We find a qualitative agreement between the two sets of results.

We can now understand better the results for
staggered hopping and uniform hopping shown in Figs.~\ref{fig14} and \ref{fig15}. Naively, one should expect that the spreading 
of the wave function should diminish as the value $V_0$ of the quasiperiodic 
potential increases. However, in the uniform hopping case and in the presence of driving, we observe in Figs.~\ref{fig14} and \ref{fig15} that the ballistic spreading velocity first increases and then decreases (Fig.~\ref{fig14}) 
and vice versa (Fig.~\ref{fig15}) with increasing $V_0$ 
depending on the values of $\omega$. It is difficult to obtain 
any analytical insight for the first case (Fig.~\ref{fig14}) since the parameter values do not allow us to perform a perturbative expansion. The second case (Fig.~\ref{fig15}) can, however, be explained 
using the Floquet van Vleck Hamiltonian described above. 
The zeroth-order term in this Hamiltonian is just the Aubry-Andr\'e 
model as expected. But the second-order term contains a second nearest-neighbor quasiperiodic hopping as well as a quasiperiodic on-site potential. Furthermore, both these terms depend on $V^{2}_{0}/\omega$. We believe that 
these competing terms in $H_F$ can give rise to a non-monotonic dependence of
the velocity of spreading as a function of $V_{0}$. To support this statement, 
we compare the wave packet dynamics obtained from the exact numerical calculation with the one obtained from the van Vleck Hamiltonian in 
Fig.~\ref{fig15}. We observe that the two results qualitatively agree 
with each other.

\section{Discussion}
\label{sec5}

We will now summarize our results. We considered a one-dimensional model
with time-independent staggered hopping amplitudes $\ga_1$ and $\ga_2$
and a periodically driven uniform hopping amplitude between
nearest-neighbor sites, and a quasiperiodic potential with strength
$V_0$. We have studied the effects of both sinusoidal driving and driving by 
a periodic square pulse. While we have mainly studied the case of
staggered hopping amplitudes, $\ga_2 = - \ga_1$, we have also looked at the 
case with uniform hopping amplitudes, $\ga_2 = \ga_1$, for comparison. 
(In the absence of driving, both these cases reduce to the Aubry-Andr\'e
model, but we find some significant differences between the two cases when
they are driven).
In the limit that the driving amplitude $a$ and frequency $\om$ are much 
larger than $\ga_1, ~\ga_2$ and $V_0$, we analytically derived a 
Floquet Hamiltonian $H_F$ to first-order in Floquet perturbation theory.
We then numerically computed the Floquet operator $U$. The 
eigenstates and eigenvalues of $U$ were found, and various properties 
of the Floquet eigenstates were then examined as follows.

We studied a generalized IPR, called $I_m^{(q)}$, of the Floquet eigenstates 
(labeled as $m$), and found the exponent of their scaling 
with the system size $L$, namely, $I_m^{(q)} \sim 1/L^{\eta_q}$. 
For standard extended states we expect $\eta_q = q-1$ for all $q$,
while standard localized states have $\eta_q = 0$ for all $q$.
We found that while most states are either extended or localized in a
standard way, there are some multifractal states which show an intermediate
behavior with $0 < \eta_q < q-1$ (we have looked at the cases $q=2, ~3$ and 4).
These states typically appear along with re-entrant transitions at intermediate
frequencies $\om$ and large values of $V_0$. A study of the average
Shannon entropy also suggests that there can be different kinds of states 
which have very different degrees of localization. In addition, we find 
re-entrant transitions between regions with extended, localized and multifractal 
states as $\om$ is varied. We note that re-entrant transitions have been 
studied earlier in periodically driven non-Hermitian systems with 
quasiperiodic potentials~\cite{longwen}.


Interestingly, we find that in the high-frequency limit 
where $\om$ is much larger than all the other parameters,
$\ga_1, ~\ga_2, V_0$ and $a$, the quasienergies are almost
the same as the energies for the undriven system ($a=0$), but the 
corresponding IPRs can have quite different values when $V_0$ is large
and $\ga_2 = - \ga_1$. Namely, when $V_0$ is 
larger than the critical value $2 \ga_1$, we know that all the states 
of the undriven system are localized. We then find that periodic driving 
of this system at a high frequency does not change the spectrum of 
quasienergies noticeably but significantly reduces some of the IPR values. 
Thus driving seems to convert some of the localized states to extended 
states. The exact mechanism by which this happens may be an interesting
problem for future studies.

In general, our numerical results show that there are many gaps in 
the quasienergy which separate states with significantly different values 
of the IPR (see Figs.~\ref{fig01} (e) and \ref{fig02} (e) in particular). 
These mobility gaps appear due to the presence of the quasiperiodic potential.
The number of gaps increases as the driving frequency is 
increased from intermediate to large values, as we see in Fig.~\ref{fig12}.
It would be useful to study exactly why this happens. Interestingly, our 
numerical results show only mobility gaps. We have not found
any examples of mobility edges which appear when there is a continuous range 
of quasienergies, and somewhere within that range there is a particular 
quasienergy which separates localized and extended states.

As a dynamical signature of localization, we have studied how a one-particle
state initialized at one particular site of the system evolves in time.
For all the parameter values that we have examined, we find that the 
mean squared displacement always increases linearly in time. The ballistic 
velocity, given by the slope of the displacement versus time, depends on the 
system parameters. The velocity generally decreases as either $\om$ or $V_0$
is increased. However, for a system with uniform hopping amplitudes 
($\ga_2 = \ga_1$) and intermediate values of $\om$, the velocity shows a 
non-monotonic dependence on $V_0$.

There is some evidence of re-entrant transitions in a recent experiment 
in which ultracold bosonic atoms are trapped in an optical lattice and the 
atoms experience a quasiperiodic potential whose strength is given periodic
$\de$-function kicks~\cite{grover}. The experiment finds regimes of
parameter space where there are multiple transitions between localized,
extended and multifractal states as the strength of the kicked quasiperiodic 
potential is varied.

\vspace{1cm}
\centerline{\bf Acknowledgments}
\vspace{.5cm}

S.A. thanks MHRD, India for financial support through a PMRF.
K.S. thanks DST, India for support through SERB project JCB/2021/000030.
D.S. thanks SERB, India for funding through Project No. JBR/2020/000043.

\end{document}